\documentclass[11pt]{article}
\usepackage[utf8]{inputenc}
\usepackage[T1]{fontenc}
\usepackage{hyperref}
\usepackage{amsmath,amssymb}
\usepackage[active]{srcltx}
\usepackage[all]{xy}
\input diagxy
\usepackage{xcolor}
\font\fr=eufm10 scaled \magstep 1 
\newtheorem{teor}{Theorem}
\newtheorem{prop}{Proposition}

\newtheorem{definition}{Definition}

\newtheorem{remark}{Remark}

\def\beq{\begin{equation}}
\def\eeq{\end{equation}}
\def\bea{\begin{eqnarray}}
\def\eea{\end{eqnarray}}
\def\beann{\begin{eqnarray*}}
\def\eeann{\end{eqnarray*}}
\def\beasn{\begin{sneqnarray}}
\def\eeasn{\end{sneqnarray}}
\def\ben{\begin{enumerate}}
\def\een{\end{enumerate}}
\def\bit{\begin{itemize}}
\def\eit{\end{itemize}}
\def\proof{ (\emph{Proof\/}) }
\newcommand{\ds}{\displaystyle}
\def\derpar#1#2{\frac{\partial{#1}}{\partial{#2}}}
\def\derpars#1#2#3{\displaystyle\frac{\partial^2{#1}}{\partial{#2}\partial{#3}}}

\def\qed{\ifvmode\Realemovelastskip\fi
{\unskip\nobreak\hfil\penalty50\hbox{}\nobreak\hfil \hbox{\vrule
height1.2ex width1.2ex}\parfillskip=0pt \finalhyphendemerits=0
\par\smallskip}}

\def\vf{\mbox{\fr X}}
\def\df{{\mit\Omega}}
\def\Lag{{\cal L}}
\def\L{{\cal L}}

\def\d{{\rm d}}
\def\H{{\cal H}}
\def\W{{\cal W}}

\def\Nat{\mathbb{N}}

\def\Real{\mathbb{R}}
\def\R{\mathbb{R}}

\newcommand{\Reeb}{\mathcal{R}}
\def\Tan{{\rm T}}
\def\Lie{\mathop{\rm L}\nolimits}
\def\inn{\mathop{i}\nolimits}
\def\Cinfty{{\rm C}^\infty}

\newcommand*{\dd}{\mathrm{d}}
\newcommand*{\contr}{\iota}
\newcommand*{\Leg}{\mathrm{Leg}}

\title{HIGHER-ORDER CONTACT MECHANICS}
\author{\sffamily 
\sc $^a$Manuel de Le\'on
\thanks{mdeleon@icmat.es\,({\it ORCID}:\,0000-0002-8028-2348).}
$^b$Jordi Gaset
\thanks{jordi.gaset@uab.cat\,({\it ORCID}:\,0000-0001-8796-3149).}
$^c$Manuel La\'inz
\thanks{manuel.lainz@icmat.es\,({\it ORCID}:\,0000-0002-2368-5853).} \\
\sc $^d$Miguel C. Mu\~noz-Lecanda
\thanks{miguel.carlos.munoz@upc.edu\,({\it ORCID}:\,0000-0002-7037-0248).}
$^d$Narciso Rom\'an-Roy
\thanks{narciso.roman@upc.edu\,({\it ORCID}:\,0000-0003-3663-9861).}
\\[1ex]
\normalsize\itshape\sffamily 
$^a$Instituto de Ciencias Matem\'aticas,
Consejo Superior de Investigaciones Cient\'ificas\\
\normalsize\itshape\sffamily 
and Real Academia de Ciencias, Madrid, Spain.
\\[1ex]
\normalsize\itshape\sffamily 
$^b$Department of Physics,
Universitat Aut\`onoma de Barcelona,
Bellaterra, Spain.
\\[1ex]
\normalsize\itshape\sffamily 
$^c$Instituto de Ciencias Matem\'aticas,
Consejo Superior de Investigaciones Cient\'ificas, Madrid, Spain.
\\[1ex]
\normalsize\itshape\sffamily 
$^d$Department of Mathematics,
Universitat Polit\`ecnica de Catalunya,
Barcelona, Spain.
}

\begin{document}

\maketitle

\pagestyle{myheadings}

\thispagestyle{empty}

\begin{abstract}
We present a complete theory of higher-order autonomous contact mechanics,
which allows us to describe higher-order dynamical systems with dissipation.
The essential tools for the theory are the extended higher-order tangent bundles,
$\Tan^kQ\times\R$, whose geometric structures are previously introduced 
in order to state the Lagrangian and Hamiltonian formalisms 
for these kinds of systems, including their variational formulation.
The variational principle, the contact forms,
and the geometric dynamical equations are obtained by using those structures 
and generalizing the  standard formulation 
of contact Lagrangian and Hamiltonian systems.
As an alternative approach, we develop a unified description that encompasses 
the Lagrangian and Hamiltonian equations as well as their relationship 
through the Legendre map; all of them are obtained 
from the contact dynamical equations and the constraint algorithm
that is implemented because, in this formalism, 
the dynamical systems are always singular.
Some interesting examples are finally analyzed using these geometric formulations.
\end{abstract}


\bigskip

\vbox{\raggedleft AMS s.\,c.\,(2020): 
{\it Primary\/}: 37J55, 53D10, 70G75, 70H50. \\
{\it Secondary\/}: 37J05, 70G45, 70H03, 70H05.}\null

\markright{{\rm M. de Le\'on} {\it et al\/},
    {\sl Higher-order contact mechanics.}}

\newpage

\medskip
\setcounter{tocdepth}{2}
{
\small
\parskip 0pt plus 0.1mm
\tableofcontents
}


\section{Introduction}

The Hamiltonian formulation for systems described by 
higher-order regular Lagrangians was first developed by Ostrogradskii in 1850, 
providing the corresponding Euler-Lagrange and Hamilton equations for the so-called higher-order mechanics.
Furthermore, the geometrization of (first-order) classical mechanics 
occurred a century later, introducing the nowadays called  {\sl symplectic mechanics},
on the cotangent bundle $\Tan^*Q$ of the configuration manifold $Q$
\cite{AM-fm, Ar,LM-sgam}. 
At about the same time, the Lagrangian geometric description was produced, 
using the almost tangent geometry of the tangent bundle $\Tan Q$
\cite{Cr-83,CP-adg,SCC-84}. However, it took a long time for the symplectic description
of higher-order Lagrangians \cite{proc:Cantrijn_Crampin_Sarlet86,LR-85,book:DeLeon_Rodrigues85,art:Gracia_Pons_Roman91,art:Gracia_Pons_Roman92}.
 
In his original paper \cite{Os-50}, M.V. Ostrogradskii stated that a theory including 
higher-order derivatives should be trivial and do not include new ideas for physics, or fatal,
since it will include additional ghost-like degrees of freedom.
However, higher-order mechanics is not a simple mathematical extension, 
and there are many cases with physical relevance
that include second (or higher) order derivatives. 
Indeed, after its geometric description, a lot of work has been developed
to understand well and apply these theories. 
A relevant point that helped a good deal was the theoretical body 
on the geometry of higher-order tangent bundles. Indeed, if the Lagrangian 
$L$ depends on derivatives up to order $k$, then 
it is defined on the tangent bundle of order $k$, $\Tan^kQ$, and the dynamics is developed on the tangent bundle of order $2k-1$,
$\Tan^{2k-1}Q$, giving the Euler-Lagrange equations of order $2k$.
In addition, the Hamiltonian description takes place on the cotangent bundle $\Tan^*(\Tan^{k-1}Q)$, and 
(for regular Lagrangians) both are connected by the {\sl Legendre map}
$\Leg\colon\Tan^{2k-1}Q \longrightarrow\Tan^*(\Tan^{k-1}Q)$.

In this paper we consider another kind of higher-order Lagrangian and Hamiltonian systems. 
The main difference is that we are concerned not only with derivatives of higher order, 
but also with an extra parameter, $z$, which 
has a different meaning as a kind of {\sl dissipation parameter}. 
So, the Lagrangian function is now defined
on the extended higher-order tangent bundle $\Tan^kQ \times \mathbb{R}$. 
First of all, we were obliged to extend the Herglotz variational principle for higher order Lagrangians, aiming to know the correct equations of motion in this situation,
which live on $\Tan^{2k-1}Q \times \mathbb{R}$.
Furthermore, the Lagrangian allows us to define a contact structure on
$\Tan^{2k-1}Q \times \mathbb{R}$ which gives the corresponding dynamics. 
Let us remark two important and non-trivial  issues: 
(1) in order  to define the contact form we are obliged to introduce a differential operator which is analogous to the total derivative
defined by W.M. Tulczyjew for the usual case; and (2) we use the contact Hamiltonian theory.
So, we obtain an analog of the Euler-Lagrange vector fields, which are semisprays of type 1, whose
integral curves are just the solutions to the motion equations for $L$.
An important remark is that, in these semisprays, the Lagrangian appears explicitly as the component in $z$.

Contact geometry \cite{BHD-2016,BGG-2017,CNY-2013,Geiges-2008} is a topic of great interest
which is being used to describe mechanical dissipative systems
\cite{CG-2019,Galley-2013,MPR-2018,Ra-2006},
both in the Hamiltonian and Lagrangian descriptions 
\cite{Bravetti2017,BCT-2017,BLMP-2020,CIAGLIA2018,DeLeon2019,DeLeon2019b,DeLeon2016b,GGMRR-2019b,LL-2018}; 
as well as field theories with dissipation \cite{GGMRR-2019,GGMRR-2020}, 
and other types of physical systems \cite{Bravetti-2019,LLM-2020,Goto-2016,KA-2013,RMS-2017}). 
Indeed, contact Hamiltonian mechanics is related with the work of G. Herglotz 
almost 90 years ago \cite{He-1930,Her-1985},
who used a generalization of the well-known Hamilton principle 
(that includes to solve an implicit differential equation before to define the action) 
that, almost miraculously, provides the same equations 
that we can obtain using contact geometry. 
For the present paper, we also have a similar situation, 
and the equations obtained by the generalized Herglotz principle 
coincide with the ones given by the geometric approach, 
showing that both approaches lead to the same result.

We also include an alternative but equivalent geometric formulation of higher-order contact systems 
which is an extension of the so-called {\sl unified Lagrangian–Hamiltonian approach}
to mechanics, and is especially useful to discuss dynamical systems described 
by degenerate Lagrangians. It is based in the original work 
by R. Skinner and R. Rusk \cite{SR-83} for first-order autonomous systems. 
Later on, this framework was generalized in order 
to describe many different kinds of systems; namely: 
time-dependent dynamical systems \cite{BEMMR-2008,CMC-2002,GM-05}, 
vakonomic and nonholonomic mechanics \cite{CLMM-2002}, 
higher-order autonomous and non-autonomous mechanical systems
\cite{art:Gracia_Pons_Roman91,
art:Gracia_Pons_Roman92,art:Prieto_Roman11,
art:Prieto_Roman12}, 
control systems \cite{BEMMR-2007,art:Colombo_Martin_Zuccalli10},
and first and higher-order classical field theories 
\cite{art:Campos_DeLeon_Martin_Vankerschaver09,LMM-2003,
ELMMR-04,PR-2015,RRS-2005,RRSV-2011,art:Vitagliano10}. 
The method that we use for this unified approach is based on the one exposed
in~\cite{LGMMR-2020} for first-order contact dynamical systems.
The advantage of this procedure is that,
in order to construct the formalism in the {\sl higher-order extended unified bundle}
$\W=\Tan^{2k-1}Q\times_Q\Tan^*(\Tan^{2k-1}Q)\times\R$,
it is not necessary to use any of the canonical structures of the higher-order tangent bundles,
but only the natural contact structure of the bundle
$\Tan^*(\Tan^{k-1}Q)\times\R$ and the natural coupling between
elements of the bundles $\Tan(\Tan^{k-1}Q)\times\R$ and $\Tan^*(\Tan^{k-1}Q)\times\R$.
Then, the contact dynamical equations in $\W$ gives both
the Euler-Lagrange and the Hamilton equations and,
being the dynamical system singular, the Legendre map and even 
the extension of the Tulczyjew total derivative
are obtained as a result of the constraint algorithm.

The paper is structured as follows:
Section \ref{prel} is devoted to review the geometric structures of higher-order tangent bundles and the basic foundations on Hamiltonian contact mechanics.
The main results of the paper are presented in Sections \ref{varform}, \ref{hocsys}, and \ref{uf}.
In particular, in Section \ref{varform} we present the variational formulation
which leads to obtain the dynamical equations for higher-order contact systems that we model geometrically in the following sections.
Thus, in Section \ref{hocsys} we define and study geometrically higher-order contact Lagrangian systems
and the corresponding Hamiltonian formalism; while
Section \ref{uf} is devoted to introduce the unified Lagrangian--Hamiltonian description
of higher-order contact systems and how the standard Lagrangian and Hamiltonian formalisms described in Section \ref{hocsys} are recovered.
Finally, in Section \ref{uex}, some interesting examples are analyzed:
first, a model of the {\sl Pais--Uhlenbeck oscillator} with dissipation,
a higher-order description for a {\sl  radiating electron},
and an academical example of a singular dissipative system.

All the manifolds are real, second countable and $\Cinfty$. 
The maps are assumed to be $\Cinfty$. Sum over repeated indices is understood.


\section{Preliminary concepts}
\label{prel}

\subsection{Geometric structures in higher-order tangent bundles}
\label{higherbundles}

(See \cite{proc:Cantrijn_Crampin_Sarlet86,book:DeLeon_Rodrigues85,art:Gracia_Pons_Roman91,art:Prieto_Roman11,book:Saunders89} for details).

{\bf Higher-order tangent bundles.}

For $k\in\Nat$,
the {\sl $k$\,th-order tangent bundle} of an $n$-dimensional manifold $Q$, 
denoted by $\Tan^kQ$,  is the $(k+1)n$-dimensional manifold
made of the $k$-jets with source at $0 \in \Real$ and target $Q$; 
that is, $\Tan^kQ = J_0^k(\Real,Q)$.
There are the canonical projections, for $1\leq r\leq k$,
$$
\begin{array}{rcclcrccl}
\rho^k_r \colon & \Tan^kQ & \longrightarrow & \Tan^rQ & , &
\rho^k_0 \colon  & \Tan^kQ & \longrightarrow & Q \\
\ & {\bf\bar c}^k(0) & \longmapsto & {\bf\bar c}^r(0) & , &
\ & {\bf\bar c}^k(0) & \longmapsto & {\bf c}(0)  \ ,
\end{array}
$$
Points of $\Tan^kQ$ are equivalence classes of curves
${\bf c}\colon I\subset\R\to Q$ by the following $k$-jet equivalence relation:
if $(U,\varphi)$ is a local chart in $Q$, with 
$\varphi=(\varphi^i)$, $1\leq i\leq n$,
and ${\bf c}(0)\in U$;
by writing $c^i=\varphi^i\circ{\bf c}$, 
the $k$-jet ${\bf\bar c}^k(0)\in\left(\rho^k_0\right)^{-1}(U)=\Tan^kU$ 
is given by $(q^i,q^i_{1},\ldots,q^i_k)$, where
$q^i=c^i(0)$ and $\ds q_\alpha^i=\frac{d^\alpha c^i}{dt^\alpha}(0)$, ($1\leq\alpha\leq k$).
If $(q^i)$ are local coordinates in $Q$, then natural coordinates in $\Tan^kQ$
are denoted $(q_0^i,q^i_1,\ldots,q^i_k)$, and local coordinates in $\Tan(\Tan^kQ)$ 
are $(q_0^i,q_1^i,\ldots,q_k^i;v_0^i,v_1^i,\ldots,v_k^i)$; 
where we have made the identification $q^i\equiv q_0^i$.
The local expressions of the above projections and their corresponding tangent maps
$\Tan\rho^k_r\colon\Tan(\Tan^kQ)\to\Tan(\Tan^rQ)$ are
\begin{eqnarray*}
\rho^k_r(q_0^i,q_1^i,\ldots,q_k^i)&=&(q_0^i,q_1^i,\ldots,q_r^i)
\\
\Tan\rho^k_r\left(q_0^i,q_1^i,\ldots,q_k^i,v_0^i,v_1^i,\ldots,v_k^i\right) &=&
\left(q_0^i,q_1^i,\ldots,q_r^i,v_0^i,v_1^i,\ldots,v_r^i\right) \ .
\end{eqnarray*}

Given a curve ${\bf c}\colon \R \to Q$,
the {\sl canonical lifting} of ${\bf c}$ to $\Tan^kQ$
is the curve ${\bf \bar c}^k\colon \Real\to\Tan^kQ$ defined as
${\bf \bar c}^k(t) = {\bf c}^k_t(0)$, where ${\bf c}_t(s)={\bf c}(s+t)$,
(the $k$-jet lifting of ${\bf c}$).
If $k=1$, we write ${\bf \bar c}^1\equiv{\bf c}'$.
In coordinates, if ${\bf c}(t)=(c^i(t))$, then
\begin{equation}
{\bf\bar c}^k=\left(c^i,\frac{d c^i}{dt},\ldots,\frac{d^kc^i}{dt^k}\right) \ .
\label{kcurve}
\end{equation}

Let $V(\rho^k_{r-1})$ be the vertical subbundle of $\Tan(\Tan^kQ)$. 
In the above coordinates, for every
 ${\rm p} \in \Tan^kQ$ and ${\rm u}\in V_p(\rho^k_{r-1})$, we have that
${\rm u}=(0,\ldots,0,v_r^i,\ldots,v_k^i)$.
We have the {\sl canonical embeddings} 
$i_{k-r+1} \colon V(\rho^k_{r-1})\hookrightarrow \Tan(\Tan^kQ)$
locally given by
$$
i_{k-r+1}(q_0^i,\ldots,q_k^i,v_r^i,\ldots,v_k^i)=
(q_0^i,\ldots,q_k^i,0,\ldots,0,v_r^i,\ldots,v_k^i) \ ,
$$
and the bundle morphisms
 $s_{k-r+1} \colon \Tan(\Tan^kQ) \to \Tan^kQ \times_{\Tan^{r-1}Q}\Tan(\Tan^{r-1}Q)$,
defined by $s_{k-r+1}({\rm u}):=\left(\tau_{\Tan^kQ}({\rm u}), \Tan\rho^k_{r-1}({\rm u})\right)$,
for every ${\rm u}\in \Tan(\Tan^kQ)$.
Their local expressions are
$$
s_{k-r+1}\left(q_0^A,\ldots,q_k^A,v_0^A,\ldots,v_k^A\right) = \left(q_0^A,\ldots,q_{r-1}^A,q_{r}^A,\ldots,q_{k}^A,v_0^A,\ldots,v_{r-1}^A\right)\ .
$$
Then we can construct the fundamental exact sequence of vector bundles
\begin{equation}
0 \longrightarrow V(\rho^k_{r-1})  \stackrel{i_{k-r+1}}{\longrightarrow}
\Tan(\Tan^kQ)  \stackrel{s_{k-r+1}}{\longrightarrow} 
\Tan^kQ \times_{\Tan^{r-1}Q}\Tan(\Tan^{r-1}Q) \longrightarrow 0
\quad , \quad r<k \ ,
\label{exseq}
\end{equation}
which, in local coordinates, is given by
\begin{align*}
0 \longmapsto & (q_0^i,\ldots,q_k^i,v_r^i,\ldots,v_k^i)\longmapsto
 (q_0^i,\ldots,q_k^i,0,\ldots,0,v_r^i,\ldots,v_k^i) \\
&(q_0^i,\ldots,q_k^i,v_0^i,\ldots,v_k^i)\longmapsto
(q_0^i,\ldots,q_{r-1}^i;q_0^i,\ldots,q_k^i;v_0^i,\ldots,v_{r-1}^i)
 \longmapsto 0 \ .
\end{align*}
Furthermore, we have the {\sl connecting maps}
$h_{k-r+1} \colon \Tan^kQ \times_{\Tan^{k-r}Q}\Tan(\Tan^{k-r}Q) \longrightarrow V(\rho^k_{r-1})$
locally defined as
$$
h_{k-r+1}\left(q_0^i,\ldots,q_{k}^i,v_0^i,\ldots,v_{k-r}^i\right) = 
\left(q_0^i,\ldots,q_k^i, 0,\ldots,0,\frac{r!}{0!}v_0^i,\frac{(r+1)!}{1!}v_1^i,\ldots,\frac{k!}{(k-r)!}v_{k-r}^i\right) \ ,
$$
which are globally well-defined and are vector bundle isomorphisms over $\Tan^kQ$.
Finally, the {\sl canonical injections} are the maps
$j_r \colon\Tan^kQ \longrightarrow \Tan(\Tan^{r-1}Q)$
defined by $j_r({\bf\bar c}^k(0)):=\mbox{\boldmath $\gamma'$}(0)$,
for ${\bf\bar c}^k(0)\in \Tan^kQ$; where
$\mbox{\boldmath $\gamma'$}\colon I\subset\Real\to\Tan(\Tan^{r-1}Q)$
is the canonical lift of the curve 
$\mbox{\boldmath $\gamma$}(t)={\bf\bar c}^{r-1}(t)$ to $\Tan(\Tan^{r-1}Q)$.
It is locally given by
$$
j_r(q_0^i,\ldots,q_k^i)=(q_0^i,\ldots,q_{r-1}^i;q_1^i,q_2^i,\ldots,q_r^i) \ .
$$

{\bf Canonical vector fields, vertical endomorphisms and almost-tangent structures.}

The following composition allows us to define the vector fields $\Delta_r\in\vf (\Tan^kQ)$,
for $1 \leq r \leq k$,
$$
\xymatrix{
\Tan^kQ \ar[rr]^-{{\rm Id}\times j_{k-r+1}} \ar@/_1.5pc/[rrrrrr]_{\Delta_r} & \ & 
\Tan^kQ \times_{\Tan^{k-r}Q} \Tan(\Tan^{k-r}Q) \ar[rr]^-{h_{k-r+1}} & \ &
 V(\rho^k_{r-1}) \ar[rr]^-{i_{k-r+1}} & \ & \Tan(\Tan^kQ)
}
$$
such that
$\Delta_r\left( q_0^i,\ldots,q_k^i \right) = 
\left( q_0^i,\ldots,q_k^i,0,\ldots,0,r!\,q_1^i,(r+1)!\,q_2^i,\ldots,\frac{k!}{(k-r)!}q_{k-r+1}^i \right)$;
that is
\begin{equation}
\label{delta-r}
\Delta_r=
\sum_{\alpha=0}^{k-r}\frac{(r+\alpha)!}{\alpha!}q_{\alpha+1}^i\derpar{}{q_{r+\alpha}^i}=
r!\,q_1^i\derpar{}{q_r^i} + 
(r+1)!\,q_2^i\derpar{}{q_{r+1}^i} + \ldots + \frac{k!}{(k-r)!}\,q_{k-r+1}^i\derpar{}{q_k^i} \ .
\end{equation}
These are the {\sl $r$\,th-canonical vector fields} in $\Tan^kQ$ and,
in particular,
\begin{equation}
\label{delta-1}
\Delta_1 = 
\sum_{\alpha=1}^{k}\alpha q_\alpha^i\derpar{}{q_\alpha^i}=
q_1^i\derpar{}{q_1^i} + 2q_2^i\derpar{}{q_2^i} + \ldots + kq_{k}^i\derpar{}{q_k^i}
\end{equation}
is the so-called {\sl Liouville vector field} in $\Tan^kQ$
and generalizes the Liouville vector field in $\Tan Q$.

In the same way, bearing in mind \eqref{exseq},
we can state the compositions
$$
\xymatrix{
\Tan(\Tan^kQ) \ar[rr]^-{s_r} \ar@/_1.5pc/[rrrrrr]_{J_r} & \ & 
\Tan^kQ \times_{\Tan^{k-r}Q} \Tan(\Tan^{k-r}Q) \ar[rr]^-{h_{k-r+1}} & \ & 
V(\rho^k_{r-1}) \ar[rr]^-{i_{k-r+1}} & \ & \Tan(\Tan^kQ) \ ;
}
$$
the map $J_r$ is called the {\sl $r$\,th-vertical endomorphism} of $\Tan(\Tan^kQ)$,
and it is locally given by
$$
J_r\left(q_0^i,\ldots,q_k^i,v_0^i,\ldots,v_k^i\right) =
 \left(q_0^i,\ldots,q_k^i,0,\ldots,0,r!\,v_0^i,(r+1)!\,v_1^i,\ldots,\frac{k!}{(k-r)!}\,v_{k-r}^i\right) \ ;
$$
that is,
\begin{equation}
\label{J-r}
J_r=\sum_{\alpha=0}^{k-r}\frac{(r+\alpha)!}{\alpha!}
\,\d q_\alpha^i\otimes\derpar{}{q_{r+\alpha}^i} \ ,
\end{equation}
(It has constant rank equal to $(k-r+1)n$).
In particular we have
\begin{equation}
\label{J-1}
J_1=\sum_{\alpha=0}^{k-1}(\alpha+1)\,\d q_\alpha^i\otimes\derpar{}{q_{\alpha+1}^i} \ ,
\end{equation}
and any other vertical endomorphism $J_r$ is obtained 
by composing $J_1$ with itself $r$ times.
As a consequence, $J_1$ defines an almost-tangent structure
of order $k$ in $\Tan^kQ$, which is called the
{\sl canonical almost-tangent structure} of $\Tan^kQ$.
We denote by $J_r^*$ the dual maps of $J_r$
(which are endomorphisms in $\Tan^*(\Tan^kQ)$)
and their natural extensions to the exterior algebra $\bigwedge(\Tan^*(\Tan^kQ))$.
Their action on differential forms is
$$
J_r^*\omega(X_1,\ldots,X_p) := \omega(J_r(X_1),\ldots,J_r(X_p)) \ ; \
\mbox{\rm for $\omega\in\df^p(\Tan^kQ)$, $X_1,\ldots,X_p \in \vf(\Tan^kQ)$} \ ,
$$
and for $f\in\Cinfty(\Tan^kQ)$ is $J_r^*(f) = f$.
The endomorphism
$J_r^* \colon \df(\Tan^kQ) \to \df(\Tan^kQ)$, $1\leq r \leq k$, 
is called the {\sl $r$\,th-vertical operator}, and it is locally given by
\begin{align*}
J_r^*(f) = f \ , \ \mbox{\rm for} \ f \in \Cinfty(\Tan^kQ) \quad ; \quad
J_r^*(\d q_\alpha^i) = \begin{cases} 0, & \mbox{if }\alpha < r \\ 
\ds \frac{\alpha!}{(\alpha-r)!}\, \d q_{\alpha-r}^i, & \mbox{if }\alpha\geq r
 \end{cases} \ .
\end{align*}
Finally, the {\sl inner contraction} of $J_r$ with
$\omega\in\df^p(\Tan^kQ)$ is the form $\inn(J_r)\omega\in\df^p(\Tan^kQ)$ defined by
$$
\inn(J_r)\omega(X_1,\ldots,X_p):= \sum_{i=1}^{p} \omega(X_1,\ldots,J_r(X_i),\ldots,X_p)\ ; \
\mbox{\rm for $X_1,\ldots,X_p \in \vf(\Tan^kQ)$} \ ,
$$
and $\inn(J_r)f = 0$, for $f\in\Cinfty(\Tan^kQ)$. 
This map $\omega\mapsto\inn (J_r)\omega$
is a derivation of degree $0$ in the $\R$-algebra of differential forms $\df(\Tan^kQ)$, which is called the
{\sl $r$\,th-vertical derivation},
whose local expression is
$$
\inn(J_r)(\d q_\alpha^i)=
\begin{cases} 0, & \mbox{\rm if }\alpha<r \\ \ds \frac{\alpha!}{(\alpha-r)!}\,\d q_{\alpha-r}^i, 
& \mbox{\rm if }\alpha\geq r \end{cases} \ .
$$

{\bf Higher-order semisprays.}

\begin{definition}
\label{semispray}
A \textbf{semispray of type $r$}, $1 \leq r \leq k$,
is a vector field $X\in\vf(\Tan^kQ)$ such that,
for every integral curve $\mbox{\boldmath $\sigma$}$ of $X$, we have that, if
 $\mbox{\boldmath $\gamma$}=\rho^k_0 \circ\mbox{\boldmath $\sigma$}$, then
 $\mbox{\boldmath $\bar\gamma$}^{k-r+1} = \rho^k_{k-r+1}\circ\mbox{\boldmath $\sigma$}$
(where $\mbox{\boldmath $\bar\gamma$}^{k-r+1}$
is the canonical lifting of $\mbox{\boldmath $\gamma$}$  to $\Tan^{k-r+1}Q$).
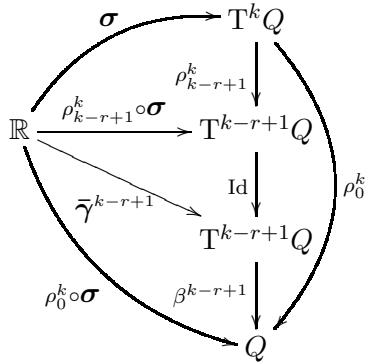
$$
\xymatrix{
\ & \ & \Tan^kQ \ar[d]_{\rho^k_{k-r+1}} \ar@/^2.5pc/[ddd]^{\rho^k_0} \\
\R \ar@/^1.5pc/[urr]^{\mbox{\boldmath $\sigma$}} \ar@/_1.5pc/[ddrr]_{\rho^k_0\circ\mbox{\boldmath $\sigma$}}
 \ar[rr]^-{\rho^k_{k-r+1}\circ\mbox{\boldmath $\sigma$}} \ar[drr]_{\mbox{\boldmath $\bar\gamma$}^{k-r+1}}
 & \ & \Tan^{k-r+1}Q \ar[d]_{{\rm Id}} \\
\ & \ & \Tan^{k-r+1}Q \ar[d]_{\beta^{k-r+1}} \\
\ & \ & Q
}
$$
In particular, $X\in\vf(\Tan^kQ)$ is a \textbf{semispray of type $1$}
or a \textbf{holonomic vector field}  in $\Tan^kQ$
if, for every integral curve $\sigma$ of $X$, we have that, if
 $\mbox{\boldmath $\gamma$}=\rho^k_0 \circ\mbox{\boldmath $\sigma$}$, then
 $\mbox{\boldmath $\bar\gamma$}^k=\mbox{\boldmath $\sigma$}$.
\end{definition}

So, semisprays of type $1$ in $\Tan^kQ$ are the vector fields 
whose integral curves are the canonical lifts to $\Tan^kQ$ of curves on $Q$
(observe that, if $k=1$, $r=1$, then $\rho^1_{1-1+1} = {\rm Id}_{\Tan Q}$,
and we recover the definition of the holonomic or {\sc sode}  vector fields in $\Tan Q$).
Thus semisprays of type $r<k$ have integral curves which are only ``partially holonomic''.

Obviously, every semispray of type $r$ is of type $s\geq r$.
We have the following equivalences for a vector field $X\in\vf(\Tan^kQ)$:

1. $X$ is a semispray of type $r$.

2. $\Tan\rho^k_{k-r} \circ X = j_{k-r+1}$

3. $J_r(X) = \Delta_r$.

The local expression of a semispray of type $r$ is
\begin{equation}
\label{semisprayk}
X = q_1^i\derpar{}{q_0^i} + q_2^i\derpar{}{q_1^i} + \ldots + q_{k-r+1}^i\derpar{}{q_{k-r}^i} + 
X_{k-r+1}^i\derpar{}{q_{k-r+1}^i} + \ldots + X_k^i\derpar{}{q_k^i} \ ,
\end{equation}
and, if $\mbox{\boldmath $\bar\gamma$}$ is an integral curve of $X$, then its components
verify the following system of differential equations of order $k+1$:
\begin{equation}
\frac{d^{k-r+2}\sigma^i}{dt^{k-r+2}} = X_{k-r+1}^i\left(\sigma^j,\frac{d\sigma^j}{dt},\ldots,\frac{d^k\sigma^j}{dt^k}\right)
 \ ,\quad  \ldots \quad , \
\frac{d^{k+1}\sigma^i}{dt^{k+1}} = X_k^i\left(\sigma^j,\frac{d \sigma^j}{dt},\ldots,\frac{d^k\sigma^j}{dt^k}\right) \ .
\label{kordersytem}
\end{equation}
For semisprays of type 1 we have
\begin{equation}
X = q_1^i\derpar{}{q_0^i} + q_2^i\derpar{}{q_1^i} + \ldots + q_k^i\derpar{}{q_{k-1}^i} + X_k^i\derpar{}{q_k^i}\ ,
\label{semisp1}
\end{equation}
and its integral curves are solutions to an ordinary differential equation of order $k+1$.

{\bf Tulczyjew's total derivative.}

The so-called {\sl Tulczyjew's total derivative} 
can be formally defined as a derivation along curves and its canonical lifts 
in a higher-order tangent bundle
\cite{book:DeLeon_Rodrigues85,art:Tulczyjew75_1}. 
Nevertheless, the simplest way to define and understand it 
is giving the following equivalent definition:

\begin{definition}
\label{Tulder}
The \textbf{Tulczyjew's total derivative}
is the map $d_T\colon \Cinfty(\Tan^rQ) \to \Cinfty(\Tan^{r+1}Q)$
which, for every $f \in \Cinfty(\Tan^rQ)$, is defined by
$$
d_Tf:=\Lie(X)(\rho^{r+1}_r)^*f  \ ,
$$
for any semispray of type 1, $X\in\vf(\Tan^{r+1}Q)$.
\end{definition}

Bearing in mind \eqref{semisp1}, we obtain the coordinate expression
$$
d_Tf=\sum_{\alpha=0}^{r}q_{\alpha+1}^i\derpar{f}{q_\alpha^i}\ .
$$
Observe that this definition is independent on the choice of the semispray of type 1,
$X\in\vf(\Tan^{r+1}Q)$, since $f=f(q_0,\ldots,q_r)$.

As it is defined as a Lie derivative, this map $d_T$ extends in a natural way
to a derivation of degree $0$ on the differential forms and,
as $d_T\d = \d d_T$, it is determined by its action on functions
and by the property
$d_T(\d q_\alpha^i) = \d q_{\alpha+1}^i$.


{\bf Higher-order Lagrangian systems}

Let $L\in\Cinfty(\Tan^kQ)$ (a {\sl Lagrangian function of order $k$}). 
Consider the Tulczyjew's total derivative
$d_T\colon \df^1(\Tan^rQ) \to \df^1(\Tan^{r+1}Q)$,
and denote $d_T^\alpha=\overbrace{d_T\circ\ldots\circ d_T}^\alpha$. 
Then, the {\sl Lagrangian $1$-form}
associated with $L$ is the semibasic form of type $k$ in  $\Tan^{2k-1}Q$ defined as
$$
\theta_L:= \sum_{\beta=1}^k (-1)^{\beta-1} \frac{1}{\beta!} d_T^{\beta-1}\inn(J_\beta)\d L \in \df^1(\Tan^{2k-1}Q)  \ ,
$$
and the corresponding {\rm Lagrangian $2$-form} is
$\omega_L:= -\d \theta_L\in \df^2(\Tan^{2k-1}Q)$.
Moreover,  the {\sl Lagrangian energy} associated with $L$
is the function defined as
$$
E_L:= \Big(\sum_{\beta=1}^k (-1)^{\beta-1} \frac{1}{\beta!} d_T^{\beta-1}(\Delta_\beta (L))\Big)-(\rho_k^{2k-1})^*L\in \Cinfty(\Tan^{2k-1}Q) \ .
$$
(It is usual to write $L$ instead of $(\rho_k^{2k-1})^*L$,
and we will do this in the sequel).

The coordinate expressions of these elements are
\begin{eqnarray*}
\theta_L &=& \sum_{\beta=1}^k \sum_{\alpha=0}^{k-\beta}(-1)^\alpha d_T^\alpha\Big(\derpar{L}{q_{\beta+\alpha}^i}\Big) \d q_{\beta-1}^i \ , \\
E_L &=& \sum_{\beta=1}^{k} \sum_{\alpha=0}^{k-\beta}  q_{\beta}^i (-1)^\alpha d_T^\alpha\left( \derpar{L}{q_{\beta+\alpha}^i} \right)-L \ .
\end{eqnarray*}
A Lagrangian function $L \in \Cinfty(\Tan^kQ)$ is said to be {\sl regular}
if $\omega_L$ is a symplectic form. Otherwise $L$ is a {\sl singular} Lagrangian.
To say that $L$ is a regular Lagrangian is locally equivalent to saying that
the Hessian matrix
$\ds \left(\frac{\partial^2L}{\partial q_k^\beta\partial q_k^\alpha}\right)$ is regular at every point of $\Tan^kQ$.

The couple $(\Tan^{2k-1}Q,\Lag)$ is said to be a {\sl Lagrangian system of order $k$},
and  the dynamical trajectories of the system are the integral curves of any
vector field $X_L \in \vf(\Tan^{2k-1}Q)$ such that
it s a solution to equation
\begin{equation}\label{eq:HOsymplectic}
\inn(X_L)\omega_L = \d E_L \ .
\end{equation}
A vector field $X_\Lag$ solution to this equation (if it exists)
is called a {\sl (higher-order) Lagrangian vector field} and
if, in addition, it is a semispray of type $1$, then it is called a
{\sl (higher-order) Euler-Lagrange vector field}, and its integral curves on the base $Q$ are
solutions to the {\sl higher-order Euler-Lagrange equations}.

\begin{remark}{\rm
The Euler-Lagrange equations can also be obtained 
introducing the \textbf{Lagrange differential}~\cite{art:Tulczyjew75_1} 
$\delta\colon\df(\Tan^kQ)\to\df(\Tan^{k+1}Q)$,
which is defined as
\begin{equation}
\label{lagrange_differential}
    \delta =\left( \sum_{\alpha = 0}^\infty \frac{(-1)^\alpha}{\alpha!} d_T^\alpha\inn(J_\alpha)\right)\circ\d \ ,
\end{equation}
where we consider that $\inn(J_\alpha)(g) = 0$ if $g \in \df(\Tan^k Q)$,
with $\alpha > k$, and we denote $J_0={\rm Id}$. 
Hence, the infinite series defining $\delta g$ has only finitely many non-zero terms.
Notice that, given a Lagrangian $L\colon\Tan^k Q \to \Real$, we have
$$
\delta L = 
\sum_{\alpha = 0}^k {(-1)}^\alpha d_T^\alpha \left(\frac{\partial L}{\partial q^i_\alpha} \right) \d q^i_\alpha \ ,
$$
and then $\delta L=0$ provides the Euler-Lagrange equations.
}
\end{remark}

\subsection{Notions on contact geometry and contact Hamiltonian systems}
\label{Hpsys}

(See, for instance,
 \cite{BCT-2017,CIAGLIA2018,DeLeon2019,GGMRR-2019b,Go-69,LL-2018} 
for details).
			
\begin{definition}
\label{dfn-contact-manifold}
A \textbf{contact manifold} is a couple $(M,\eta)$,
where $M$ is a $(2n+1)$-dimensional manifold and
$\eta\in\df^1(M)$ is a differential $1$-form
such that $\eta\wedge(\d\eta)^{\wedge n}$ is a volume form in $M$.
The form $\eta$ is called a \textbf{contact form} in $M$.
\end{definition}

As a consequence of this definition, for every contact manifold
we have the decomposition
$\Tan M=\ker\d\eta\oplus\ker\eta\equiv\mathcal{D}^{\rm R}\oplus\mathcal{D}^{\rm C}$. 
Then,
if $(M,\eta)$ is a contact manifold, there exists a unique vector field $\Reeb\in\vf(M)$ 
such that
\begin{equation}\label{eq-Reeb}
        \inn(\Reeb)\d\eta = 0\quad ,\quad
        \inn(\Reeb)\eta = 1.
\end{equation}
It is called the {\sl Reeb vector field} and generates ${\cal D}^{\rm R}$
which is known as the {\sl Reeb distribution}.

The manifold $\Tan^*Q\times\R$ is
the canonical model for contact manifolds since,
if $\theta_0\in\df^1(\Tan^*Q)$ 
and $\omega_0=-\d\theta_0\in\df^2(\Tan^*Q)$
are the canonical forms in $\Tan^*Q$,
$z$ is the Cartesian coordinate of $\R$, and 
$\pi_1\colon \Tan^*Q \times\R \to \Tan^*Q$ 
is the canonical projection, then 
$\eta=\d z-\pi_1^*\theta_0$ is a contact form in  $\Tan^*Q\times\R$,
and the Reeb vector field is
$\displaystyle\Reeb= \frac{\partial}{\partial z}$.
Furthermore, if $(M,\eta)$ is a contact manifold,
for every point $p\in M$, there exists a chart 
$(U; q^i, p_i, z)$, $1\leq i\leq n$, such that
$$
\eta\vert_U= \d z - p_i\,\d q^i \quad ; \quad
\Reeb\vert_U= \frac{\partial}{\partial z}\  .
$$
These are the \textsl{Darboux} or \textsl{canonical coordinates} 
of the contact manifold $(M,\eta)$.

If $(M,\eta)$ is a contact manifold and $H\in\Cinfty(M)$,
then there exists a unique vector field $X_H\in\vf(M)$ such that
    \begin{equation}
\label{hamilton-contact-eqs}
            \inn(X_H)\d\eta=\d H-({\Reeb}(H))\eta
\quad ,\quad
            \inn(X_H)\eta=-H \ ;
    \end{equation}
and the integral curves ${\bf c}\colon I\subset\R\to M$ of $X_H$
are the solutions to equations
    \begin{equation}\label{hamilton-contactc-curves-eqs}
            \inn({\bf c}')\d\eta=\left(\d H-({\Reeb}(H))\eta\right)\circ{\bf c}
\quad ,\quad
            \inn({\bf c}')\eta=-H\circ{\bf c} \ ,
    \end{equation}
    where ${\bf c}'\colon I\subset\R\to \Tan M$ is the canonical lift of
    ${\bf c}$ to $\Tan M$.

\begin{definition}
The vector field $X_H$ is said to be the
\textbf{contact Hamiltonian vector field} associated to $H$ and equations 
\eqref{hamilton-contact-eqs} and \eqref{hamilton-contactc-curves-eqs}
are the \textbf{contact Hamiltonian equations} 
for $X_H$ and its integral curves, respectively.
Then, the triple $(M,\eta,H)$ is a \textbf{contact Hamiltonian system}.
\end{definition}

\begin{remark}
{\rm The contact Hamiltonian equations \eqref{hamilton-contact-eqs}  can be equivalently written as
\begin{equation*}
 \Lie({X_H})\eta= -(\Reeb(H)) \, \eta \quad ,\quad  \inn(X_H)\eta = -H \ .
\end{equation*}
In addition, given a contact manifold $(M,\eta)$,  there exists the $\Cinfty(M)$-module isomorphism
$$
\begin{array}{rccl}
   \flat\colon & \vf(M) & \longrightarrow & \df^1(M) \\
   & X & \longmapsto & \inn(X)\d\eta+(\inn(X)\eta)\eta
\end{array} \ ,
$$
and \eqref{hamilton-contact-eqs} are also equivalent to
$$
\flat(X_H) =\d H - (\Reeb(H) + H) \eta \ .
$$
Finally, equations \eqref{hamilton-contact-eqs} can also be written  without making use of the Reeb vector field $\Reeb$ since,
if we take $U=\{p\in M;H(p)\not= 0\}$ and $\Omega = -H\,\d\eta + \d H\wedge\eta$ on~$U$,
the contact Hamiltonian vector field $X_H\in\vf(U)$ is the solution to
$$
\inn(X_H)\Omega = 0 \quad,\quad \inn(X_H)\eta = -H\ ,
$$
and its integral curves ${\bf c}\colon I\subset\R\to M$  are solutions to
$$
\inn({\bf c}')\Omega = 0 \quad , \quad\inn({\bf c}')\eta = - H\circ{\bf c} \:. 
$$
}\end{remark}

In Darboux coordinates $(q^i,p_i,z)$,
$$ 
X_H = \frac{\partial H}{\partial p_i}\frac{\partial}{\partial q^i} - 
\left(\frac{\partial H}{\partial q^i} + 
p_i\frac{\partial H}{\partial z}\right)\frac{\partial}{\partial p_i} + 
\left(p_i\frac{\partial H}{\partial p_i} - H\right)\frac{\partial}{\partial z}\ ; 
$$
and its integral curves ${\bf c}(t) = (q^i(t), p_i(t), z(t))$
are solutions to equations \eqref{hamilton-contactc-curves-eqs} which read as
$$
    \dot q^i = \frac{\partial H}{\partial p_i}\quad ,\quad
    \dot p_i = -\left(\frac{\partial H}{\partial q^i} + p_i\frac{\partial H}{\partial z}\right)\quad ,\quad
    \dot z = p_i\frac{\partial H}{\partial p_i} - H\ .
$$

Unlike in symplectic Hamiltonian systems, the energy and the geometric structure of contact Hamiltonian systems are not conserved. Indeed,
$$
 \Lie (X_H) H  = - \Reeb(H)\,H
\quad , \quad \Lie (X_H) \eta  = - \Reeb(H)\,\eta \ ,
$$
or, what is equivalent,
\begin{equation}
\label{contact_dissipation}
({\phi^{X_H}_t})^*H  = \sigma^H_t \,H \quad, \quad  ({\phi^{X_H}_t})^*\eta  = \sigma^H_t \,\eta \ ,
\end{equation}
where $\phi^{X_H}$ is the flow of $X_H$, and $\sigma^H_t \in\Cinfty(M)$ 
is the so-called {\sl dissipation rate} of the system.

\begin{remark} {\rm
If some of the conditions stated in Definition 
\ref{dfn-contact-manifold} do not hold, then
$(M,\eta)$ is said to be a {\sl precontact manifold}~\cite{DeLeon2019}.
In this case, Reeb vector fields are not uniquely defined and the map $\flat$ is not an isomorphism.
}\end{remark}


\section{Higher-order Herglotz variational principle}
\label{varform}

The higher-order contact dynamical equations,
that allow us to describe geometrically higher-order dynamical systems with dissipation, are derived from a variational principle
which is stated as an extension of the classical {\sl Herglotz variational principle} for first-order dynamical systems \cite{He-1930,Her-1985}.
In this Section we state this variational formulation,
as a previous step and a justification of the geometrical setting of the
equations that are presented and applied in the following sections.

If ${\rm q}_{0},{\rm q}_{1} \in\Tan^k Q$, with components $({q}^i_{0,\alpha}), ({q}^i_{1,\alpha})$,
we denote by  $\Omega({\rm q}_0,{\rm q}_1)$ the space of smooth curves $\mathbf{c}:[0,1] \to Q$ 
such that their $k$-jet lifts fulfil 
${\mathbf{\bar c}}^k_\alpha(0)=q_{0,\alpha},{\mathbf{ \bar c}}^k_\alpha(1)=q_{1,\alpha}$,
for $0\leq\alpha\leq k$. 
The $k$\,th tangent space of $\Omega({\rm q}_0, {\rm q}_1)$ at a curve $\mathbf{c}$
is given by sections of the $k$\,th tangent bundle along $\mathbf{c}$ 
that vanishes at the endpoints; that is,
$$
    \begin{aligned}
        \Tan_\mathbf{c} \Omega({\rm q}_0, {\rm q}_1) =  \{ &
            \delta {\bf c} = (\mathbf{c}; \delta \mathbf{c}_0, \delta \mathbf{c}_1, \ldots, \delta \mathbf{c}_{k-1}) :[0,1] \to \Tan^k Q \mid  \\ &
            \rho^k_0 \circ \delta \mathbf{c} =  \mathbf{c},
            \delta \mathbf{c}_\alpha(0)= 0, \, \delta \mathbf{c}_\alpha(1)=0 \text{ for } 0\leq \alpha < k 
            \} \ .
    \end{aligned}
$$

Now, a Lagrangian function of order $k$ is just a function $L\colon\Tan^kQ\times\R\to\R$. We fix $z_0 \in \Real$.
We define the operator 
$$
    \mathcal{Z}_{L,z_0}\colon \Omega({\rm q}_0, {\rm q}_1) \to \Cinfty ([0,1]) 
$$
 which assigns to each curve $\mathbf{c}$ the function $\mathcal{Z}_{L,z_0} (\mathbf{c})$ that solves the following differential equation
\begin{equation}
\label{contact_var_0de}
\begin{cases}
   \ds \frac{\d \mathcal{Z}_{L,z_0}(\mathbf{c})}{\d t} &= L(\bar{\mathbf{c}}^k, \mathcal{Z}_{L,z_0} (\mathbf{c})),\\
    \mathcal{Z}_{L,z_0}(\mathbf{c})(0) &= z_0\ .
    \end{cases}
\end{equation}

\begin{definition}
The \textbf{higher-order contact action functional} associated with a Lagrangian function $L$ is 
the map which assigns to each curve $\mathbf{c}$, the solution to equation
\eqref{contact_var_0de} evaluated at the endpoint:
$$
 \begin{aligned}
  \mathcal{A}_{L,z_0}: \Omega({\rm q}_0, {\rm q}_1) &\to \R,\\
   c &\mapsto \mathcal{Z}_{L,z_0}(\mathbf{c})(1) \ .
 \end{aligned}
$$
\end{definition}

\begin{remark}
    The contact action $\mathcal{A}_{L,z_0}$ is a generalization of the higher-order Euler-Lagrange action. Indeed, if $L$ does not depend on $z$, we can solve the ODE~\eqref{contact_var_0de}, obtaining
    \begin{equation}
        \mathcal{A}_{L,z_0} (c) = \int_{0}^1 L(\bar{\mathbf{c}}^k(t)) \d t + z_0 \ ,
    \end{equation}
    which coincides with the Euler-Lagrange action up to the constant $z_0$.
\end{remark}

We look for the critical points of the functional $A_{L,z_0}$ and try to obtain a set of differential equations characterizing them. These equations will be called the \textbf{$(2k)$th order Herghlotz equations}.. The proof is a generalization of the one presented in~\cite{DeLeon2019} for first-order Lagrangians.

\begin{teor}[Higher-order Herglotz variational principle]
Let $L\colon\Tan^k Q \times \R \to \R$ be a Lagrangian function and let 
$\mathbf{c}\in \Omega({\rm q}_0, {\rm q}_1)$ and $z_0 \in \R$. 
Then, $\mathbf{c}$ is a critical point of $\mathcal{A}_{L,z_0}$  if, and only if, 
$(\bar{\mathbf{c}}, \mathcal{Z}_{L,z_0}(\mathbf{c}))$ satisfies the following equations
\begin{equation}
\label{herglotz1}
   \sum_{\alpha = 0}^k {(-1)}^\alpha
\mathcal{D}_{L,z_0,\mathbf{c}}^\alpha \left(\frac{\partial L}{\partial {q}_\alpha^i}(\bar{\mathbf{c}}^k(t),\mathcal{Z}_{L,z_0}(\mathbf{c})(t)) 
    \right) = 0 \ ;
\end{equation}
where, for $f\colon\R\to\R$,  the operator $\mathcal{D}_{L,z_0,\mathbf{c}}$ is defined as
\begin{equation}
\label{Dderivative}
(\mathcal{D}_{L,z_0,\mathbf{c}} f) (t) := \frac{1}{\sigma_{L,z_0}} \frac{\d (\sigma_{L,z_0} f)}{\d t } (t)=
 \frac{\d f}{\d t}(t) - \frac{\partial L}{\partial z}(\bar{\mathbf{c}}^k(t),\mathcal{Z}_{L,z_0}(\mathbf{c})(t)) f (t)\ ;
\end{equation}
and  $\sigma_{L,z_0}\colon\R\to\R$ is the solution to the Cauchy problem
$$
\begin{cases}
    \ds \frac{\d \sigma_{L,z_0}}{\d t} &= \ds -\frac{\partial L}{\partial z} \sigma_{L,z_0} \ ,\\
\sigma_{L,z_0}(0) &= z_0 \ .
\end{cases}
$$
Equations~\eqref{herglotz1} are called the {\rm $(2k)$\,th order Herglotz equations}.
\label{Herglotzteor}
\end{teor}
\proof
    Let $\mathbf{c} \in \Omega({\rm q}_0,{\rm q}_1)$. Then  $\mathbf{c}$ is a critical curve of the map  $\mathcal{A}_{L,z_0}$, if and only if $(\Tan_\mathbf{c} \mathcal{A}_{L,z_0})(\delta {\bf c})=0$ for every $\delta \mathbf{c} \in\Tan_\mathbf{c} \Omega({\rm q}_0, {\rm q}_1)$. 

First we compute $ (\Tan_\mathbf{c} \mathcal{Z}_{L,z_0})(\delta {\bf c})$. 
Let $\mathbf{c}_\lambda \in \Omega({\rm q}_0,{\rm q}_1)$, where $\lambda\in(-\delta,\delta)\subset\Real$,  be a smoothly parametrized family of curves such that
   $\delta \mathbf{c}$ are their $k$-jets with respect to $\lambda$ at $\lambda=0$. That is:
   $$
   \delta \mathbf{c}=(\mathbf{c}; \delta \mathbf{c}_0,\ldots, \delta \mathbf{c}_{k-1})\quad,\quad  \delta \mathbf{c}_0=\frac{\d}{\d\lambda}|_{\lambda=0}\mathbf{c}_\lambda,\, \delta \mathbf{c}_1=\frac{\d}{\d\lambda}|_{\lambda=0}\dot{\mathbf{c}}_\lambda,\ldots,\delta \mathbf{c}_{k-1}=\frac{\d}{\d\lambda}|_{\lambda=0}\frac{\d^{k-1}}{\d t^{k-1}}\mathbf{c}_\lambda \ .
   $$
   We also denote
   \begin{equation*}
    \delta \mathbf{c}_{k}=\frac{\d}{\d\lambda}|_{\lambda=0}\frac{\d^{k}}{\d t^{k}}\mathbf{c}_\lambda \ .
   \end{equation*}
   Observe that, by direct calculus we obtain
   \begin{equation}\label{variattangent}
   \delta \mathbf{c}_{\alpha}=\frac{\d^{\alpha}}{\d t^{\alpha}}\delta \mathbf{c}_0
\quad , \quad \mbox{\rm (for $0\leq \alpha \leq k$) } \ .
   \end{equation}
    In order to simplify the notation, we write $\psi = (\Tan_\mathbf{c} \mathcal{Z}_{L,z_0})(\delta {\bf c})$. Since $\mathcal{Z}_{L,z_0} (\mathbf{c}_\lambda)(0)=z_0$ for all $\lambda$, then $\psi(0)=0$. Also notice that $\psi(1) = (\Tan_\mathbf{c} \mathcal{Z}_{L,z_0})(\delta {\bf c})(1) = (\Tan_\mathbf{c} \mathcal{A}_{L,z_0})(\delta {\bf c})$.
   We have that:
   $$
   \psi = (\Tan_\mathbf{c} \mathcal{Z}_{L,z_0})(\delta {\bf c})=\frac{\d}{\d\lambda}\mathcal{Z}_{L,z_0} (\mathbf{c}_\lambda)|_{\lambda=0}\ .
   $$ 
   Then, computing the derivative of $\psi$ with respect to $t$, interchanging the derivatives, we have:
\beann
   \dot\psi(t)&=&\frac{\d}{\d t}\frac{\d}{\d\lambda}\mathcal{Z}_{L,z_0} (\mathbf{c}_\lambda)(t)\vert_{\lambda=0}=
   \frac{\d}{\d\lambda}\vert_{\lambda=0}\left(\frac{\d}{\d t}\mathcal{Z}_{L,z_0} (\mathbf{c}_{\lambda})(t)\right)\\
   &=&\frac{\d}{\d\lambda}\vert_{\lambda=0}{L(\bar{\mathbf{c}}^k_\lambda(t),\mathcal{Z}_{L,z_0}(\mathbf{c}_\lambda)(t))}\\
   &=&
            \sum_{\alpha = 0}^{k} \frac{\partial L}{\partial q^i_\alpha}(\bar{\mathbf{c}}^k(t),\mathcal{Z}_{L,z_0}(\mathbf{c})(t)) {\delta c}_\alpha^i(t) + 
            \frac{\partial L}{\partial z}(\bar{\mathbf{c}}^k(t),\mathcal{Z}_{L,z_0}(\mathbf{c})(t)) \psi(t) \\
            &=&A(t)+B(t)\psi(t)  \ .
\eeann
   Hence $\psi$ is the solution to the initial condition problem:
   $$
   \dot\psi(t)=A(t)+B(t)\psi(t),\quad \psi(0)=0\, .
   $$  
   The solution is given by
   $$
   \psi(t)=\exp\left(\int_0^tB(s)\d s\right)\int_0^t\d u\, A(u)\exp\left(-\int_0^tB(s)\d s\right)\ ;
   $$
   and writing: 
   $$
        \sigma_{L,z_0}(t) = \exp \left({-\int_0^t \frac{\partial L}{\partial z}(\bar{\mathbf{c}}^k(\tau),\mathcal{Z}_{L,z_0}(\mathbf{c})(\tau)) \d \tau}\right) > 0 \ ,
   $$
   we have:
   $$
    \psi(t) = \frac{1}{\sigma_{L,z_0}(t)} \int_0^t \sigma_{L,z_0}(\tau) \left(
    \sum_{\alpha = 0}^{k} \frac{\partial L}{\partial q^i_\alpha} (\bar{\mathbf{c}}^k(\tau),\mathcal{Z}_{L,z_0}(\mathbf{c})(\tau)) {\delta c}_\alpha^i(\tau) \right) \d \tau \ .
   $$
   Now integrating by parts, as usually in variational calculus, using equation (\ref{variattangent}) and taking into account that variations vanish at the extreme points, we have that:
   \begin{equation*}
       \begin{split}
           \psi(1) = (\Tan_\mathbf{c} \mathcal{A}_{L,z_0})(\delta {\bf c}) & = \frac{1}{\sigma_{L,z_0}(t)}
           \int_0^1  \sum_{\alpha=0}^{k} {\delta c}_0^i(\tau) {(-1)}^\alpha 
           \frac{\d^\alpha }{\d \tau^\alpha} \left(\sigma(\tau)\frac{\partial L}{\partial {q}_\alpha^i}(\bar{\mathbf{c}}^k(\tau),\mathcal{Z}_{L,z_0}(\mathbf{c})(\tau)) 
           \right) \d \tau \ .
       \end{split}
   \end{equation*}
   We know that $\psi$ must vanish if the curve ${\mathbf c}$ must be a critical point, then as the variations $\delta\mathbf{ c}_0$ are arbitrary, by the fundamental theorem of calculus of variations, we obtain
   $$
       \frac{1}{\sigma_{L,z_0}(t)}\sum_{\alpha = 0}^{k} {(-1)}^\alpha
       \frac{\d^\alpha }{\d t^\alpha} \left(\sigma(t)\frac{\partial L}{\partial {q}_\alpha^i}(\bar{\mathbf{c}}^k(t),\mathcal{Z}_{L,z_0}(\mathbf{c})(t))
       \right) = 0 \, ,
   $$
   where we have used that $ \sigma_{L,z_0}:[0,1]\to\Real$, as it has been defined, satisfies the conditions
   $$
     \frac{\d \sigma_{L,z_0}}{\d t} =  -\frac{\partial L}{\partial z} \sigma_{L,z_0}
     \,\ ,\quad
   \sigma_{L,z_0}(0) = z_0\, .
   $$   
   Then, introducing the operator $\mathcal{D}_{L,z_0,\mathbf{c}}$ defined in \eqref{Dderivative},
   this equation reduces to~\eqref{herglotz1}.
\qed


\section{Higher-order contact systems}
\label{hocsys}

Using the geometric structures of higher-order tangent bundles and bearing in mind the setting of contact mechanics
we can establish the Lagrangian and Hamiltonian formalisms for
higher-order contact mechanics in a geometrical way.

\subsection{Geometric structures in $\Tan^kQ\times\Real$}

Consider the {\sl extended higher-order tangent bundle} $\Tan^kQ\times\R$, 
which has natural coordinates $(q_0^i,q^i_1,\ldots,q^i_k,z)$,
and the canonical projections (for $r<k$)
\begin{eqnarray*}
z\colon \Tan^kQ\times\R\to\R 
\ &,& \
\tau^k\colon \Tan^kQ\times\R\to\Tan^kQ
 \ , \\
\tau_r^k\colon \Tan^kQ\times\R\to\Tan^rQ\times\R
 \ &,& \
\tau_0^k\colon \Tan^kQ\times\R\to Q\times\R\ .
\end{eqnarray*}

We denote by $\d z$ the volume form in $\Real$, and its pull-backs
to all the manifolds.

We have that
$\Tan(\Tan^kQ \times \R) =
\Tan(\Tan^kQ)\oplus_{\Tan^kQ\times\Real}\Tan\R$;
then any operation between tangent vectors to $\Tan^kQ$
can be extended to tangent vectors to $\Tan^kQ \times\R$.
In particular, the $r$\,th- vertical endomorphisms $J_r$
of $\Tan(\Tan^kQ)$ and the $r$\,th-canonical vector fields $\Delta_r$
on $\Tan^kQ$ yield vertical endomorphisms
${\cal J}_r \colon \Tan (\Tan^kQ\times\R) \to \Tan (\Tan^kQ\times\R)$ 
and canonical vector fields which are also denoted
$\Delta_r \in \vf(\Tan Q\times\R)$
(and, in particular, the {\sl Liouville vector field} $\Delta_1$
and the {\sl almost-tangent structure} ${\cal J}_1$).
Obviously, their expressions in natural coordinates are
\eqref{delta-r}, \eqref{delta-1}, \eqref{J-r}, and \eqref{J-1}, again.

If ${\bf c} \colon\R \rightarrow Q\times\R$ 
is a curve, with ${\bf c}=(\mathbf{c}_1,\mathbf{c}_0)$;
the \textsl{prolongation} of ${\bf c}$ to $\Tan^kQ\times\R$ 
is defined as the curve
${\bf \widetilde c}^k= (\mathbf{\bar c}^k_1,\mathbf{c}_0)
\colon \R \longrightarrow \Tan^kQ \times \R$,
where $\mathbf{\bar c}_1^k$ is the prolongation of ${\bf c}_1$ to $\Tan^kQ$.
Then ${\bf \widetilde c}$ is said to be a \textsl{holonomic} curve in $\Tan^kQ\times\R$.
In coordinates, if ${\bf c}(t)=(q^i(t), z(t))$,
bearing in mind \eqref{kcurve}, then
$\ds {\bf \widetilde c}^k=\left(q^i(t),\frac{d q^i}{dt}(t),\ldots,\frac{d^kq^i}{dt^k}(t),z(t)\right)$.

\begin{definition}
A vector field $\Gamma\in\vf(\Tan^kQ\times\R)$ 
is a \textbf{semispray of order $r$ in $\Tan^kQ\times\R$}
if its integral curves 
$\mbox{\boldmath$\sigma$}=({\bf\bar c}^r,{\bf c}_0)\colon\R\to\Tan^kQ\times\R$
are such that the curves ${\bf\bar c}^r\colon\R\to\Tan^kQ$
satisfy the conditions stated in Definition \ref{semispray}.
In particular $\Gamma$ is a \textbf{semispray of order $1$} or a
\textbf{holonomic vector field  in $\Tan^kQ\times\R$}
if its integral curves are holonomic.
\end{definition}

Obviously, $\Gamma\in\vf(\Tan^kQ\times\R)$
is holonomic if, and only if,
${\cal J}_1(\Gamma)=\Delta_1$.

Taking into account \eqref{semisprayk}, 
the local expression of a semispray of order $r$ in $\Tan^kQ\times\R$ is
$$
\Gamma= 
q_1^i\derpar{}{q_0^i}+q_2^i\derpar{}{q_1^i}+\ldots
+q_{k-r+1}^i\derpar{}{q_{k-r}^i}+ 
X_{k-r+1}^i\derpar{}{q_{k-r+1}^i}+\ldots + X_k^i\derpar{}{q_k^i}+ 
g\,\frac{\partial}{\partial z}
\ ,
$$
and its integral curves are solutions to the system of
differential equations \eqref{kordersytem} together with
$\ds \frac{dz}{dt}=g$,
with $X^i_\alpha,g\in \Cinfty(\Tan^kQ\times\R)$.
In particular, for semisprays of type 1 we have
$$
\Gamma= 
q_1^i\derpar{}{q_0^i}+q_2^i\derpar{}{q_1^i}+\ldots
+q_k^i\derpar{}{q_{k}^i} + X_k^i\derpar{}{q_k^i}+ 
g\,\frac{\partial}{\partial z} \ .
$$

\subsection{Higher-order contact Lagrangian systems}
\label{sec-conLagsys}

Let $ L\colon\Tan^kQ\times\R\to\R$.
We call it a \textsl{Lagrangian function of order $k$}.

In order to define a suitable contact Lagrangian form associated with
a higher-order Lagrangian function,
first we need to define the analogous to the total derivative $d_T$
in the extended higher-order tangent bundles $\Tan^rQ\times\R$.
Unlike what happens in $\Tan^rQ$, where the total derivative is canonically defined
using only semisprays of type 1 (see Definition \ref{Tulder}),
 this is not so in the present situation and, 
as it is justified in the above Section \ref{varform}, 
the corresponding operator is associated with the Lagrangian function.

Thus, if $L\in\Cinfty(\Tan^kQ\times\R)$,
for $k\leq r<2k-1$, we consider the family of vector fields
$\vf_1^L(\Tan^rQ\times\R)\subset\vf(\Tan^rQ\times\R)$
made of the vector fields $\Gamma\in \vf(\Tan^rQ\times\R)$ such that:

(i) they are semisprays of type 1 in $\vf(\Tan^rQ\times\R)$,

(ii) the condition $\inn(\Gamma)\d z=L$ holds.

\noindent 
Their coordinate expressions are
$$
\Gamma= 
q_1^i\derpar{}{q_0^i}+q_2^i\derpar{}{q_1^i}+\ldots
+q_r^i\derpar{}{q_{r-1}^i} + X_r^i\derpar{}{q_r^i}+L\derpar{}{z} \ .
$$

\begin{definition} 
We introduce the operator
$d_L\colon \Cinfty(\Tan^rQ\times\R) \to \Cinfty(\Tan^{r+1}Q\times\R)$,
$k\leq r<2k-1$,
which, for every $F \in \Cinfty(\Tan^rQ\times\R)$, is defined by
$$
d_LF:=\Lie(\Gamma)(\tau^{r+1}_r)^*F \quad , \quad \mbox{\rm for any $\Gamma\in \vf_1^L(\Tan^{r+1}Q\times\R)$} \ .
$$
Then, the \textbf{Lagrangian total derivative}
is the map $D_L\colon \Cinfty(\Tan^rQ\times\R) \to \Cinfty(\Tan^{r+1}Q\times\R)$
which, for every $F \in \Cinfty(\Tan^rQ\times\R)$, is defined by
$$
D_LF:=d_LF-\derpar{L}{z}(\tau^{r+1}_r)^*F \ .
$$
\end{definition}

(We have denoted by $L$ the pull-back of $L$ to any higher-order bundle
by the corresponding projection.
We will continue to do so in the sequel).

Note that $D_L$ is well-defined since $\ds\derpar{}{z}$ is a canonical
vector field in $\Tan^rQ\times\R$,
and that this definition is independent on the choice of
$\Gamma\in\vf_1^L(\Tan^{r+1}Q\times\R)$.
The coordinate expression is
\begin{equation}
\label{DL_operator}
D_L F:=\sum_{\alpha=0}^{r}q_{\alpha+1}^i\derpar{F}{q_\alpha^i}+L\derpar{F}{z}-\derpar{L}{z} F \ .
\end{equation}

\begin{remark}{\rm
This Lagrangian total derivative $D_L$ is closely related to the operator
$\mathcal{D}_{L,z_0,\mathbf{c}}$ introduced in \eqref{Dderivative}.
In fact; in Theorem \ref{Herglotzteor}, $\mathcal{D}_{L,z_0,\mathbf{c}}$ is a differential operator acting on functions of a real variable, since all
the functions are evaluated along curves and their canonical liftings.
In this way, if $F\in\Cinfty(\Tan^kQ\times\R)$, 
we can retrieve $D_L F\in\Cinfty(\Tan^{k+1}Q\times\R)$, as it is given  in~\eqref{DL_operator} as follows:
\begin{gather*}
\mathcal{D}_{L,z_0,\mathbf{c}} (F(\bar{\mathbf{c}}^k,\mathcal{Z}_{L,z_0}(\mathbf{c})) )=
\frac{d}{dt}F(\bar{\mathbf{c}}^k,\mathcal{Z}_{L,z_0}(\mathbf{c}))
-\Big(F\derpar{L}{z}\Big)(\bar{\mathbf{c}}^k,\mathcal{Z}_{L,z_0}(\mathbf{c})) \\ = 
\sum_{\alpha=0}^{k}c_{\alpha+1}^i\derpar{F}{q_\alpha^i} (\bar{\mathbf{c}}^k,\mathcal{Z}_{L,z_0}(\mathbf{c}) )+\left(L\derpar{F}{z}\right)(\bar{\mathbf{c}}^k,\mathcal{Z}_{L,z_0}(\mathbf{c}))-\left(\derpar{L}{z}F\right)(\bar{\mathbf{c}}^k,\mathcal{Z}_{L,z_0}(\mathbf{c})) \ ;
\end{gather*}
    that is,
$$
(D_L F) (\bar{\mathbf{c}}^k,\mathcal{Z}_{L,z_0}(\mathbf{c})) = \mathcal{D}_{L,z_0,\mathbf{c}} (F(\bar{\mathbf{c}}^k,\mathcal{Z}_{L,z_0}(\mathbf{c})) \ .
$$
This fact justifies the definition given for $D_L$.
}
\end{remark}

\begin{remark}{\rm
Observe that $d_L$ is a derivation but $D_L$ is not; indeed,
$$
D_L (f g) = D_L(f) g + f D_L (g) + fg \frac{\partial L}{\partial z} \ .
$$
}
\end{remark}

\begin{remark}{\rm
The maps $d_L$ and $D_L$ extend in a natural way
to differential forms,
$d_L,D_L\colon \df^p(\Tan^rQ\times\R) \to \df^p(\Tan^{r+1}Q\times\R)$, for every $p$;
 and,  for $\xi \in \df^p(\Tan^rQ\times\R)$, 
we have that
$$
D_L\xi:=d_L\xi
-\derpar{L}{z}(\tau^{r+1}_r)^*\xi
:=\Lie(\Gamma)(\tau^{r+1}_r)^*\xi
-\derpar{L}{z}(\tau^{r+1}_r)^*\xi 
\quad , \quad \mbox{\rm for any $\Gamma\in\vf_1^L(\Tan^{r+1}Q\times\R)$} \ .
$$
}
\end{remark}

Using this Lagrangian total derivative we can define:

\begin{definition}
\label{lagrangean}
The \textbf{Lagrangian energy}
associated with $ L$ is the function 
$$
E_L =\sum_{\beta=1}^k (-1)^{\beta-1}\frac{1}{\beta!} D_L^{\beta-1}(\Delta_\beta( L))-L\in\Cinfty(\Tan^{2k-1}Q\times\R) \ .
$$

The \textbf{Cartan Lagrangian forms} associated with $ L$ are
\begin{eqnarray*}
\theta_L &=&  \sum_{\beta=1}^k (-1)^{\beta-1}\frac{1}{\beta!}
 D_L^{\beta-1} \inn({\cal J}_\beta)d L
\in\df^1(\Tan^{2k-1}Q\times\R)   \ .
\\
\omega_L &=& -\d \theta_L =
\sum_{\beta=1}^k (-1)^\beta \frac{1}{\beta!} \d  D_L^{\beta-1}\inn({\cal J}_\beta)d L
\in\df^2(\Tan^{2k-1}Q\times\R) \ .
\end{eqnarray*}

The \textbf{contact Lagrangian form} associated with $L$ is
$$
\eta_L=\d z-\theta_L\in\df^1(\Tan^{2k-1}Q\times\R) \ .
$$

The couple $(\Tan^{2k-1}Q\times\R,L)$, endowed with these structures,
 is  a \textbf{(pre)contact Lagrangian system (of order $k$)}.
\end{definition}

The coordinate expressions of these elements are
\begin{eqnarray}
\label{Energy}
E_L &=&
\sum_{\beta=1}^{k}\sum_{\alpha=0}^{k-\beta}  q_{\beta}^i (-1)^\alpha D_L^\alpha\left( \derpar{L}{q_{\beta+\alpha}^i} \right)-L  \ ,  \\
\label{Lag1Form}
\theta_L &=& 
\sum_{\beta=1}^k \sum_{\alpha=0}^{k-\beta}(-1)^\alpha D_L^\alpha\left(\derpar{L}{q_{\beta+\alpha}^i}\right) \d q_{\beta-1}^i
 \ , \\
\label{Lag2Form}
\omega_L &=&
\sum_{\beta=1}^k \sum_{\alpha=0}^{k-\beta}(-1)^{\alpha+1}\d\,D_L^\alpha\left(\derpar{L}{q_{\beta+\alpha}^i}\right) \wedge \d q_{\beta-1}^i
 \ ,  \\
\label{ContactForm}
\eta_L &=& 
\d z-\sum_{\beta=1}^k \sum_{\alpha=0}^{k-\beta}(-1)^\alpha D_L^\alpha\left(\derpar{L}{q_{\beta+\alpha}^i}\right) \d q_{\beta-1}^i \ . 
\end{eqnarray}

Consider the bundle $ \Tan^*(\Tan^{k-1}Q)\times\R$ and the natural projections
\begin{eqnarray*}
z\colon \Tan^*(\Tan^{k-1}Q)\times\R\to\R  \quad , &
\pi_1\colon \Tan^*(\Tan^{k-1}Q)\times\R\to \Tan^*(\Tan^{k-1}Q) \ , \\
&\pi_{\Tan^{k-1}Q} \colon \Tan^*(\Tan^{k-1}Q) \to \Tan^{k-1}Q \ .
\end{eqnarray*}

\begin{definition}
Let $L\in\Cinfty(\Tan^kQ\times\R)$ be a Lagrangian function.
The \textbf{extended Legendre---Ostrogradskii map} associated with $L$
is the map $\widetilde\Leg \colon\Tan^{2k-1}Q\times\R\to\Tan^*(\Tan^{k-1}Q\times\R)$
which is defined as follows 
\cite{book:DeLeon_Rodrigues85,art:Prieto_Roman12}:
if $\tau\colon\Tan(\Tan^{2k-1}Q\times\R)\to\Tan^{2k-1}Q\times\R$
is the canonical projection,
for every ${\rm  u}\in \Tan(\Tan^{2k-1}Q\times\R)$
and ${\rm p}=\tau({\rm u})$,
 $$
(\theta_L)_{\rm p}({\rm u}):= 
\langle \Tan\rho^{2k-1}_{k-1}({\rm u}),\widetilde\Leg({\rm p}) \rangle \ .
$$
Then, if $\mu\colon\Tan^*(\Tan^{k-1}Q\times\R)\to\Tan^*(\Tan^{k-1}Q)\times\R$
is the canonical projection, the   \textbf{generalized Legendre map\/}
associated with $L$ is the map
$\Leg \colon \Tan^{2k-1}Q\times\R \to \Tan^*(\Tan^{k-1}Q)\times\R$
given by 
$$\Leg:=\mu\circ\widetilde\Leg \ ; $$
that is, we have the diagram
$$
\xymatrix{
\Tan^{2k-1}Q\times\R \ar[rr]^-{\widetilde\Leg} \ar@/_1.5pc/[rrrr]_{\Leg} & \ & 
\Tan^*(\Tan^{k-1}Q\times\R) \ar[rr]^-{\mu}  & \ & \Tan^*(\Tan^{k-1}Q)\times\R} \ .
$$
\end{definition}

These maps are bundle morphisms (on the base $\Tan^{k-1}Q\times\R$).
Bearing in mind the local expression (\ref{Lag1Form})
of the form $\theta_L$, 
if $(q_0^i,q_1^i,\ldots,q_{k-1}^i,p^0_i,p^1_i,\ldots,p^{k-1}_i)$
denote the canonical coordinates in $\Tan^*(\Tan^{k-1}Q)$,
after some calculations
we obtain that the local expression of the extended Legendre map is
\begin{equation}
  \Leg^*z = z \  , \
  \Leg^*q^i _{r-1}= q^i_{r-1} \  , \
  \Leg^*p_i^{r-1}=\sum_{\alpha=0}^{k-r}(-1)^\alpha D_L^\alpha\left(\derpar{L}{q_{r+\alpha}^i}\right)\equiv\widehat p_i^{\ r-1}  \ ; \quad (1 \leq r \leq k) \ . 
\label{legmap}
\end{equation}
The functions $\widehat p_i^{\ r-1}$ are called {\sl generalized momenta} or
{\sl Jacobi-Ostrogradskii momenta};
they satisfy the relation
\begin{equation}
\widehat p^{\ r-1}_i = \derpar{L}{q_r^i} - D_L(\widehat p^{\ r}_i) \ ,
\label{recursivep}
\end{equation}
and so, all the momentum coordinates can be obtained recursively from $\widehat p_i^{\ k-1}$. 

Using the local expression of the generalized Legendre map
it is immediate to prove that:

\begin{prop}
The following conditions for a Lagrangian function $L\in\Cinfty(\Tan^kQ\times\R)$ are equivalent:
\begin{enumerate}
\item
The generalized Legendre map $\Leg$ is a local diffeomorphism.
\item
$(\Tan^{2k-1}Q\times\R,\eta_L)$ is a $(2kn+1)$-dimensional contact manifold.
\item
The Hessian matrix 
$\ds W_{ij} = 
\left( \frac{\partial^2L}{\partial q^i_k\partial q^j_k}\right)$ 
is everywhere nonsingular.
\end{enumerate}
\end{prop}

\begin{definition}
A Lagrangian  $L$ is \textbf{regular} if the above equivalent conditions hold;
otherwise $L$ is a \textbf{singular} Lagrangian.
In particular, $L$ is \textbf{hyperregular} 
if $\Leg$ is a global diffeomorphism.

A singular Lagrangian is said to be \textbf{almost-regular} if:
(i) $P_0=\Leg(\Tan^{2k-1}Q\times\R)$
is a closed submanifold of $\Tan^*(\Tan^{k-1}Q)\times\R$,
(ii) $\Leg$ is a submersion onto its image $P_0$,
and (iii) for every ${\rm p}\in\Tan^{2k-1}Q\times\R$, the fibres
 $\Leg^{-1}(\Leg({\rm p}))$
 are connected submanifolds of $\Tan^{2k-1}Q\times\R$.
\end{definition}

It is obvious that every regular contact Lagrangian system 
has associated the contact Hamiltonian system
$(\Tan^{2k-1}Q\times\R, \eta_L, E_L)$
whose \textsl{Reeb vector field}  $\Reeb_L\in\vf(\Tan^{2k-1}Q\times\R)$ 
is uniquely determined by the relations
$$
    \inn(\Reeb_L)\d\eta_L=0 \quad ,\quad
    \inn(\Reeb_L)\eta_L=1 \ ,
$$
and its local expression is
$$
\Reeb_L=\frac{\partial}{\partial z}+f^i_\alpha\frac{\partial}{\partial q^i_\alpha}
 \ , \
$$
with $f^i_\alpha=0$, for $0\leq\alpha\leq k$, and the others given by the recursion formula
$$
f^i_{\alpha}=(-1)^{\alpha-k}W^{ij}\left(\derpar{\widehat{p}^{\ i}_{2k-\alpha-1}}{z}+ 
\sum_{r=1}^{\alpha-k}f^{l}_{\alpha-r}\derpar{\widehat{p}^{\ i}_{2k-\alpha-1}}{q^{\alpha-r}_l}\right) 
\quad , \quad (k<\alpha<2k)\ ;
$$
where $(W^{ji})$ is the inverse of the Hessian matrix;
that is, $W^{ji} W_{ik} = \delta^j_k$.

Furthermore, a simple calculation shows that
$$
\Lie(\Reeb_L)E_L=-\derpar{L}{z} \ .
$$

\subsection{Higher-order contact Euler--Lagrange (Herglotz) equations}

Next we set the dynamical equations of
higher-order Lagrangian systems in a geometrical way.

\begin{definition}
\label{def-lageqs}
Let $(\Tan^{2k-1}Q\times\R,L)$ be a Lagrangian system of order $k$.

The \textbf{$2k$\,th contact Euler--Lagrange} or 
\textbf{$2k$\,th order Herglotz equations} for a holonomic curve
${\bf\widetilde c}\colon I\subset\R \to\Tan^{2k-1}Q\times\R$ are
\begin{equation}
\inn({\bf\widetilde c}')\d\eta_L = 
\Big(\d E_L - (\Reeb_L(E_L))\eta_L\Big)\circ{\bf\widetilde c} 
\quad ,\quad
\inn({\bf\widetilde c}')\eta_L = - E_L\circ{\bf\widetilde c} 
\ ,
\label{hec}
\end{equation}
where ${\bf\widetilde c}'\colon I\subset\R\to\Tan(\Tan^{2k-1}Q\times\R)$ denotes the
canonical lifting of ${\bf\widetilde c}$ to $\Tan(\Tan^{2k-1}Q\times\R)$.

The \textbf{contact Lagrangian equations} for a vector field $X_L\in\vf(\Tan^{2k-1}Q\times\R)$ are 
\begin{equation}
\label{eq-E-L-contact1}
    \inn(X_L)\d \eta_L=\d E_L-(\Reeb_L(E_L))\eta_L
\quad , \quad
    \inn(X_L)\eta_L=-E_L \ .
\end{equation}
A vector field which is a solution to these equations is called a
\textbf{(higher-order) contact Lagrangian vector field}
(it is the contact Hamiltonian vector field for the function $E_L$).
\end{definition}

\begin{remark}{\rm
We can define the morphism
\begin{eqnarray*}
        \flat_L\colon\Tan(\Tan^{2k-1}Q \times \mathbb{R}) 
        &\to&\Tan^*(\Tan^{2k-1}Q \times \mathbb{R}) \\
        v & \mapsto & \contr_{v} \dd \eta_L + \eta_L(v) \eta_L.
\end{eqnarray*}
which extends in a natural way to vector fields,
$ \flat_L\colon\vf(\Tan^{2k-1}Q \times \mathbb{R})\to\df^1(\Tan^{2k-1}Q \times \mathbb{R})$,
and hence equations \eqref{eq-E-L-contact1} are equivalent to
$$
    \flat_L(\xi_L) = \dd E_L - (\Reeb_L(E_L) + E_L) \eta_L \ .
$$
Furthermore, in $U=\{p\in M;\H(p)\not= 0\}$,
if $\Omega_L=-E_L\,\d\eta_L + \d E_L\wedge\eta_L$,
equations \eqref{hec} and \eqref{eq-E-L-contact1} are equivalent to
\begin{eqnarray*}
\inn({\bf \widetilde c}')\Omega_L = 0 & ,&
\inn({\bf \widetilde c}')\eta_L = - E_L\circ{\bf\widetilde c} \ ,
\\
\inn(X_L)\Omega_L = 0 &, &
\inn(X_L)\eta_L = -E_L\ .
\end{eqnarray*}
}\end{remark}

In natural coordinates, for a holonomic curve
$\ds{\bf \widetilde c}=\left(q^i(t),\frac{d q^i}{dt}(t),\ldots,\frac{d^kq^i}{dt^k}(t),z(t)\right)$,
the $2k$\,th-order contact Euler-Lagrange or Herglotz equations \eqref{hec} are
\begin{eqnarray}
\label{ELeqs2}
\sum_{\alpha=0}^{k}(-1)^\alpha D_L^\alpha\Big(\derpar{\Lag}{q_{\alpha}^i}\Big) ({\bf \widetilde c}(t)) &=& 0 \ , 
\\
\label{ELeqs1}
\frac{dz}{dt}&=&L \ ,
\end{eqnarray}
Observe that equation~\eqref{herglotz1} is just~\eqref{ELeqs2} evaluated along $(\bar{\mathbf{c}}^k(t),\mathcal{Z}_{L,z_0}(\mathbf{c}))$.
On the other hand, for equations \eqref{eq-E-L-contact1}, as
\begin{eqnarray*}
\d E_L &=& 
\sum_{r=1}^k \sum_{\alpha=0}^{k-r}(-1)^\alpha D_L^\alpha \left(\derpar{L}{q_{r+\alpha}^i} \right)\d q_r^i +
 \sum_{r=1}^k \sum_{i=0}^{k-r} (-1)^i\sum_{\beta=0}^{k} q_r^i D_L^i\left(\frac{\partial^2L}{\partial q_\beta^j\partial q_{r+i}^i}
 \d q_\beta^j \right) 
\\ & & 
- \sum_{r=0}^{k} \derpar{L}{q_r^i} \d q_r^i +
\left(\sum_{r=1}^{k}\sum_{\alpha=0}^{k-r} (-1)^\alpha  q_{r}^i D_L^\alpha\left( \frac{\partial^2L}{\partial z\partial q_{r+\alpha}^i} \right)-\derpar{L}{z}\right)\d z \ ,
\end{eqnarray*}
then, if
$\ds X_L = \sum_{\alpha=0}^{2k-1} f_\alpha^i \derpar{}{q_\alpha^i}+g\derpar{}{z}$, equations \eqref{eq-E-L-contact1} lead to
\begin{eqnarray*}
0&=&\Big(f_0^j-q_1^j\Big) \frac{\partial^2L}{\partial q_k^j\partial q_k^i} \ , \\
0&=&\Big(f_{1}^j-q_{2}^j\Big)\frac{\partial^2L}{\partial q_k^j\partial q_k^i}-
\Big(f_0^j-q_{1}^j\Big)(\cdots\cdots)  \ , \\
& & \qquad \qquad \qquad \vdots \\
0&=&\Big(f_{2k-2}^j - q_{2k-1}^j\Big)\frac{\partial^2L}{\partial q_k^j\partial q_k^i} -
 \sum_{\alpha=0}^{2k-3}\Big(f_\alpha^j-q_{\alpha+1}^j \Big) (\cdots\cdots) \ , \\
0&=&(-1)^k\Big(f_{2k-1}^j-D_L (q_{2k-1}^j)\Big) \frac{\partial^2L}{\partial q_k^j\partial q_k^i}+ \sum_{\alpha=0}^{k} (-1)^\alpha D_L^\alpha\Big( \derpar{L}{q_\alpha^i}\Big) - 
\sum_{\alpha=0}^{2k-2} \Big(f_\alpha^j-q_{\alpha+1}^j \Big) (\cdots\cdots)  \, , \\
0&=&g-L-\sum_{\alpha=0}^{2k-2} \left(f_\alpha^j-q_{\alpha+1}^j \right) (\cdots\cdots)  \ ,
\end{eqnarray*}
where the terms inside the brackets $(\cdots\cdots)$ contain relations involving partial derivatives
of the Lagrangian and applications of $D_L$ on them, 
which are omitted for simplicity.
Then we have:

\begin{teor}
If $L$ is a regular Lagrangian, then there exists a unique vector field 
$X_L\in\vf(\Tan^{2k-1}Q\times\R)$ solution to \eqref{eq-E-L-contact1}
and it is a semispray of order $1$; that is, a holonomic vector field in $\Tan^{2k-1}Q\times\R$,
which is called the \textbf{(higher-order) Euler--Lagrange vector field} 
associated with $L$. Its integral curves are the solutions to
equations \eqref{ELeqs2} and \eqref{ELeqs1}.
\end{teor}
\proof
If $L$ is regular then the Hessian matrix 
$\ds\left( \frac{\partial^2L}{\partial q^i_k\partial q^j_k}\right)$ is regular,
then the first $2k-1$ of the above equations give
the conditions for $X_L$ to be a semispray of order $1$ and
the two last groups of equations allow us to determine univocally
the last coefficients $f_{2k-1}^j$ and $g$ of $X_L$.
Therefore the integral curves of $X_L$ are holonomic and solutions
to  \eqref{ELeqs2} and \eqref{ELeqs1}; that is, to \eqref{hec}.
\qed

\begin{remark}{\rm
\indent If $L$ is singular, as $(\Tan^{2k-1}Q\times\R,\eta_L,E_L)$ is a precontact Lagrangian system,
solutions to the Lagrangian equations
are not necessarily holonomic vector fields in $\Tan^{2k-1}Q\times\R$ and 
the condition ${\cal J}_1(X_L)=\Delta_1$ must be added to the Lagrangian equations
in order to obtain the $2k$\,th-order contact Euler-Lagrange or Herglotz equations \eqref{hec}
for the integral curves of $X_L$.
In general, these equations are not compatible everywhere on $\Tan^{2k-1}Q\times\R$ 
and a constraint algorithm must be implemented in order to find 
a submanifold $S_f\hookrightarrow\Tan^{2k-1}Q\times\R$
(if it exists) where there are holonomic vector fields $X_L\in\vf(\Tan^{2k-1}Q\times\R)$,
tangent to $S_f$, which are solutions to the above equations on $S_f$.
Furthermore, although  Reeb vector fields are not uniquely determined
in precontact manifolds,
the constraint algorithm and the final dynamics
are independent of the Reeb. (See \cite{DeLeon2019}
for a discussion on all these topics).
}\end{remark}

\begin{remark}{\rm
 In analogy to the Lagrange differential~\eqref{lagrange_differential}, 
we can also define the \textbf{Herglotz\ differential}
$\delta_L\colon\df^p(\Tan^rQ\times\Real)\to\df^{p+1}(\Tan^{2r}Q\times\Real)$
as follows:
$$
\delta_L = \left(\sum_{\alpha = 0}^\infty \frac{(-1)^\alpha}{\alpha!} D_L^\alpha\inn({\cal J}_\alpha)\right)\circ\d \ .
$$
Again, we consider that $\inn({\cal J}_\alpha)(g) = 0$, if $g \in \df^p(\Tan^r Q \times \Real)$, 
with $\alpha > r$, and in this case we identify ${\cal J}_0$
with the natural projection $\Tan(\Tan^kQ\times \Real)$ onto $\Tan(\Tan^k Q)$.
Given a Lagrangian $L\colon\Tan ^k Q \times \Real \to \Real$, we have
$$
\delta_L L = \sum_{\alpha = 0}^k (-1)^\alpha D_L^\alpha \left(\frac{\partial L}{\partial q^i_\alpha} \right) \d q^i_\alpha \ ,
$$
and therefore $\delta_L L=0$ provides the Herglotz equations.
}
\end{remark}

\begin{remark}{\rm
The function $\sigma_{L,z_0}$ introduced in Theorem \ref{Herglotzteor}
 is related with the dissipation rate of the system 
$(T^k Q \times \Real, \eta_L, E_L)$
(as it is defined in~\eqref{contact_dissipation}) in the following way:
$$
\sigma_{L,z_0}(t) =  \sigma^{E_L}_t(\bar{\mathbf{c}}(t), \mathcal{Z}_{L,z_0}(\mathbf{c}(t))) \ .
$$
Therefore, as for first-order systems, the following relations hold:
\beq
 \Lie (X_L)E_L = - \Reeb_L(E_L) \,E_L
\quad , \quad \Lie (X_L) \eta_L  = - \Reeb_L(E_L)\, \eta_L \ ,
\label{LieEL}
\eeq
or, what is equivalent,
$$
({\phi^{X_L}_t})^*E_L  = \sigma_t^{E_L}\,E_L \quad, \quad  
({\phi^{X_L}_t})^*\eta_L = \sigma_t^{E_L}\,\eta_L \ .
$$
Notice that \eqref{LieEL} expresses the dissipation of the energy of the system.
}
\end{remark}

\begin{remark}{\rm
\indent To appreciate the structure of the higher-order contact Euler-Lagrange equations \eqref{ELeqs2} when expanded, the case $k=2$ could be illustrative. In this situation the Lagrangian depends on $(q^i_0,q^i_1,q^i_2,z)$, which includes accelerations. The associated contact structure and the Lagrangian energy are given by
\begin{eqnarray*}
\eta_L&=&\d z-\frac{\partial L}{\partial q^i_2}\d q^i_1-\left(\frac{\partial L}{\partial q^i_1}-q^j_3\frac{\partial^2 L}{\partial q^j_2\partial q^i_2}-q^j_2\frac{\partial^2 L}{\partial q^j_1\partial q^i_2}-q^j_1\frac{\partial^2 L}{\partial q^j_0\partial q^i_2}-L\frac{\partial^2 L}{\partial z\partial q^i_2}+\frac{\partial L}{\partial z}\frac{\partial L}{\partial q^i_2}\right)\d q^i_0\ ;
\\
E_L&=&q^i_2\frac{\partial L}{\partial q^i_2}+q^i_1\left(\frac{\partial L}{\partial q^i_1}-q^j_3\frac{\partial^2 L}{\partial q^j_2\partial q^i_2}-q^j_2\frac{\partial^2 L}{\partial q^j_1\partial q^i_2}-q^j_1\frac{\partial^2 L}{\partial q^j_0\partial q^i_2}-L\frac{\partial^2 L}{\partial z\partial q^i_2}+\frac{\partial L}{\partial z}\frac{\partial L}{\partial q^i_2}\right)-L\ .
\end{eqnarray*}
Hence, the second-order contact Euler-Lagrange equations for a holonomic curve are
\begin{eqnarray*}
0&=&\frac{\partial L}{\partial q^i_0}-D_L\frac{\partial L}{\partial q^i_1}+D_LD_L\frac{\partial L}{\partial q^i_2}=q^j_4\frac{\partial^2 L}{\partial q^j_2\partial q^i_2}+q^j_3q^k_3\frac{\partial^3 L}{\partial q^k_2\partial q^j_2\partial q^i_2}
\\
&+&q^k_3\biggl(-\frac{\partial^2 L}{\partial q^k_2\partial q^i_1}+\frac{\partial^2 L}{\partial q^k_1\partial q^i_2}+2q^j_2\frac{\partial^3 L}{\partial q^k_2\partial q^j_1\partial q^i_2}+2q^j_1\frac{\partial^3 L}{\partial q^k_2\partial q^j_0\partial q^i_2}+\frac{\partial L}{\partial q^k_2}\frac{\partial^2 L}{\partial z\partial q^i_2}+2L\frac{\partial^3 L}{\partial q^k_2\partial z\partial q^i_2}
\\
&-& \frac{\partial^2 L}{\partial q^k_2\partial z} \frac{\partial L}{\partial q^i_2}-2\frac{\partial L}{\partial z} \frac{\partial^2 L}{\partial q^k_2\partial q^i_2}\biggr)
\\
&+&q^k_2\biggl(-\frac{\partial^2 L}{\partial q^k_1\partial q^i_1}+q^j_2\frac{\partial^3 L}{\partial q^k_1\partial q^j_1\partial q^i_2}+\frac{\partial^2 L}{\partial q^k_0\partial q^i_2}+2q^j_1\frac{\partial^3 L}{\partial q^k_1\partial q^j_0\partial q^i_2}+\frac{\partial L}{\partial q^k_1}\frac{\partial^2 L}{\partial z\partial q^i_2}+2L\frac{\partial^3 L}{\partial q^k_1\partial z\partial q^i_2}
\\
&-& \frac{\partial^2 L}{\partial q^k_1\partial z} \frac{\partial L}{\partial q^i_2}-2\frac{\partial L}{\partial z} \frac{\partial^2 L}{\partial q^k_1\partial q^i_2 }\biggr)
\\
&+&q^k_1\biggl(-\frac{\partial^2 L}{\partial q^k_0\partial q^i_1}+q^j_1\frac{\partial^3 L}{\partial q^k_0\partial q^j_0\partial q^i_2}+\frac{\partial L}{\partial q^k_0}\frac{\partial^2 L}{\partial z\partial q^i_2}+2L\frac{\partial^3 L}{\partial q^k_0\partial z\partial q^i_2}-\frac{\partial^2 L}{\partial q^k_0\partial z} \frac{\partial L}{\partial q^i_2}-2\frac{\partial L}{\partial z} \frac{\partial^2 L}{\partial q^k_0\partial q^i_2 }\biggr)
\\
&+& L^2\frac{\partial^3 L}{\partial z\partial z\partial q^i_2}-L\frac{\partial^2 L}{\partial z \partial z} \frac{\partial L}{\partial q^i_2 }-L\frac{\partial L}{\partial z}\frac{\partial^2 L}{\partial z\partial q^i_2}- \frac{\partial L}{\partial z}\frac{\partial L}{\partial z}\frac{\partial L}{\partial q^i_2}-L\frac{\partial^2 L}{\partial z\partial q^i_1}+\frac{\partial L}{\partial z}\frac{\partial L}{\partial q^i_1}+\frac{\partial L}{\partial q^i_0} \ .
\end{eqnarray*}
}\end{remark}

\subsection{Higher-order Hamiltonian description}

First, we consider the hyperregular case (the regular case is the same, but restricting
on the suitable subsets where $\Leg$ is a local diffeomorphism).
In this case, the generalized Legendre map $\Leg$ is a
diffeomorphism between $\Tan^{2k-1}Q\times\R$ and $\Tan^*(\Tan^{k-1}Q)\times\R$.

As it was pointed out at the beginning of section \ref{Hpsys},
the bundle $\Tan^*(\Tan^{k-1}Q)\times\R$ is endowed with a
canonical contact structure which is constructed as follows:
if $\vartheta_{k-1}\in\df^1(\Tan^*(\Tan^{k-1}Q))$ and
$\varpi_{k-1}=-\d\vartheta_{k-1}\in\df^2(\Tan^*(\Tan^{k-1}Q))$
are the canonical $1$ and $2$ forms of the cotangent bundle $\Tan^*(\Tan^{k-1}Q)$;
we denote $\theta_{k-1}:=\pi_1^*\vartheta_{k-1}\in\df^1(\Tan^*(\Tan^{k-1}Q)\times\R)$
and $\omega_{k-1}:=\pi_1^*\varpi_{k-1}\in\df^2(\Tan^*(\Tan^{k-1}Q)\times\R)$;
then the canonical contact form on $T^*(T^{k-1}  Q) \times \mathbb{R}$ is
$$
\eta_{k-1} = \dd z - \theta_{k-1}\in\df^1(\Tan^*(\Tan^{k-1}Q)\times\R) \ ,
$$
and $(\Tan^*(\Tan^{k-1}Q)\times\R,\eta_{k-1})$ is a $(2kn+1)$-dimensional contact manifold
with Reeb vector field $\ds\Reeb_{k-1}=\derpar{}{z}\in\vf(\Tan^*(\Tan^{k-1}Q)\times\R)$.

Taking into account the local expressions 
\eqref{Energy}, \eqref{Lag1Form}, \eqref{Lag2Form},
and \eqref{legmap}we can write 
\begin{equation}
\theta_L = \sum_{\alpha=0}^{k-1} \widehat p^{\ \alpha}_i\d q_\alpha^i
\ , \
\omega_L = \sum_{\alpha=0}^{k-1}\d \widehat p^{\ \alpha}_i\wedge\d q_\alpha^i
\ , \
\eta_L=\d z- \sum_{\alpha=0}^{k-1} \widehat p^{\ \alpha}_i\d q_\alpha^i
\ , \
E_L = \sum_{\alpha=0}^{k-1} \widehat{p}^{\ \alpha}_i q^i_{\alpha+1} - L
\ ,
\label{coor3}
\end{equation}
and then it is obvious that
$$
\theta_L=\Leg^*\theta_{k-1} \quad , \quad
\omega_L=\Leg^*\omega_{k-1}  \quad ,\quad
\eta_L=\Leg^*\eta_{k-1}
 \ .
$$
In addition, there exists (maybe locally) a unique function 
$H\in\Cinfty(\Tan^*(\Tan^{k-1}Q)\times\R)$ such that $\Leg^*H=E_L$,
which is called the {\sl Hamiltonian function} associated to this system.
In coordinates, as
$\ds E_L = \sum_{\alpha=0}^{k-1} \widehat{p}^{\ \alpha}_i q^i_{\alpha+1} - L$,
then 
$$
H=\sum_{\alpha=0}^{k-1} p^\alpha_i q^i_{\alpha+1}-(L\circ\Leg^{-1})\ .
$$

\begin{definition}
The triad $(\Tan^*(\Tan^{k-1}Q)\times\R,\eta_{k-1},H)$ is the
\textbf{canonical contact Hamiltonian system} associated to the hyperregular Lagrangian
system $(\Tan^{2k-1}Q\times\R,L)$.
\end{definition}

Now, as a consequence of the general theory of contact Hamiltonian systems
(see Section \ref{Hpsys}) we have that:

\begin{teor}
\label{teo-hameqs}
There exists a unique vector field $X_H\in\vf(\Tan^*(\Tan^{k-1}Q)\times\R)$ such that
    \begin{equation}
\label{Hameqs}
            \inn(X_H)\d\eta_{k-1}=\d H-({\Reeb}_{k-1}(H))\eta_{k-1}
\quad , \quad
            \inn(X_H)\eta_{k-1}=-H \ .
    \end{equation}
    Then, the integral curves of $X_H$, ${\bf c}\colon I\subset\R\to\Tan^*(\Tan^{k-1}Q)\times\R$,
    are the solutions to equations
    \begin{equation}
\label{Hamcureqs}
            \inn({\bf c}')\d\eta_{k-1}=(\d H-({\Reeb}_{k-1}(H))\eta_{k-1})\circ{\bf c}
\quad , \quad
            \inn({\bf c}')\eta_{k-1}=-H\circ{\bf c} \ ,
    \end{equation}
    where ${\bf c}'\colon I\subset\R\to \Tan(\Tan^*(\Tan^{k-1}Q)\times\R)$ is the canonical lift of
    ${\bf c}$ to $\Tan(\Tan^*(\Tan^{k-1}Q)\times\R)$.
\end{teor}

\begin{definition}
The vector field $X_H$ is the
\textbf{higher-order Hamiltonian vector field} associated with $H$ and equations 
\eqref{Hameqs} and \eqref{Hamcureqs}
are the \textbf{higher-order contact Hamiltonian equations} 
for this vector field and its integral curves, respectively.
\end{definition}

In natural coordinates $(q_\alpha^i,p^\alpha_i,z)$ of $\Tan^*(\Tan^{k-1}Q)\times\R$,
($0\leq\alpha\leq k-1$; $1\leq i\leq n$), 
taking
$$
 X_H = f_\alpha^i \derpar{}{q_\alpha^i} + g^\alpha_i \derpar{}{p^\alpha_i} + g\derpar{}{z} \ ,
$$
and using the coordinate expressions \eqref{coor3}, 
from\eqref{Hameqs}we obtain that
$$
f^i_\alpha=\frac{\partial H}{\partial p_i^\alpha} \quad , \quad
g_i^\alpha=-\left( \frac{\partial H}{\partial q^i_\alpha}+p_i^\alpha\frac{\partial H}{\partial z}\right)
\quad , \quad
g= p_i^\alpha\frac{\partial H}{\partial p_i^\alpha} - H \ .
$$
Then, if ${\bf c}\colon I\subset\R\to\Tan^*(\Tan^{k-1}Q)\times\R$ is an integral curve of $X_H$,
its components $(q^i_\alpha(t),p_i^\alpha(t),z(t))$ are the solutions to equations
\begin{equation}
\frac{dq^i_\alpha}{dt}= \frac{\partial H}{\partial p_i^\alpha} \quad , \quad
\frac{dp_i^\alpha}{dt}=- \left(\frac{\partial H}{\partial q^i_\alpha}+p_i^\alpha\frac{\partial H}{\partial z}  \right)
\quad , \quad
\frac{dz}{dt} = p_i^\alpha\frac{\partial H}{\partial p_i^\alpha} - H \ .
\label{Heqcoor}
\end{equation}
which are the local expression of equations \eqref{Hamcureqs}

Finally,  using the relation $\eta_L=\Leg^*\eta_{k-1}$ and taking into account that
$\Leg$ is a diffeomorphism, it is immediate to prove that
$$\Leg_*X_L=X_H \ ,$$
which establishes the equivalence between the higher-order Lagrangian and Hamiltonian formalisms.

For singular higher-order Lagrangian systems,
in general there is no way to associate a canonical Hamiltonian formalism,
unless minimal regularity conditions are imposed \cite{art:Gracia_Pons_Roman91}.
In particular these conditions hold when the Lagrangian is almost-regular.
In this case, we have the submanifold
$j_0\colon P_0=\Leg(\Tan^{2k-1}Q\times\R)\hookrightarrow\Tan^* (\Tan^{k-1}Q)\times\R$.
We can define the form $\eta_0:=j_0^*\eta_{k-1}$
and the restriction of the extended Legendre map to this submanifold,
$\Leg_0\colon\Tan^{2k-1}Q\times\R\to P_0$, by
$\Leg = j_0\circ\Leg_0$.
Obviously we have that $\eta_L=\Leg^*\eta_{k-1}=\Leg_0^*\eta_0$.
Furthermore, the Lagrangian energy  $E_L$ is a $\Leg_0$-projectable function, 
and then there exists a unique function $H_0\in \Cinfty(P_0)$ 
such that $\Leg_0^*H_0 = E_L$ \cite{art:Gracia_Pons_Roman91}.
Then:

\begin{definition}
The triad $(P_0,\eta_0,H_0)$, where $\eta_0=j_0^*\eta_{k-1}$, is the
\textbf{canonical precontact Hamiltonian system} associated with the almost regular Lagrangian system
$(\Tan^{2k-1}Q,L)$.
\end{definition}

For this system, if $\Reeb_0\in\vf(P_0)$ is a Reeb vector field,
and for $X_{H_0}\in\vf(P_0)$, we have the contact Hamilton equations
    \begin{equation}
\label{Hameqs0}
            \inn(X_{H_0})\d\eta_0=\d H_0-({\Reeb}_0(H_0))\eta_0
\quad , \quad
            \inn(X_{H_0})\eta_0=-H_0 \ .
    \end{equation}
If $X_{H_0}$ is a solution to these equations, then its integral curves, 
${\bf c}\colon I\subset\R\to P_0$,
are the solutions to equations
    \begin{equation}
\label{Hamcureqs0}
            \inn({\bf c}')\d\eta_0=(\d H-({\Reeb}_0(H_0))\eta_0)\circ{\bf c}
\quad , \quad
            \inn({\bf c}')\eta_0=-H_0\circ{\bf c} \ ,
    \end{equation}
    where ${\bf c}'\colon I\subset\R\to \Tan P_0$ is the canonical lift of
    ${\bf c}$ to the tangent bundle $\Tan P_0$.

\begin{remark}{\rm
In general, these (pre)contact Hamiltonian equations are not compatible on $P_0$ 
and a constraint algorithm must be implemented in order to find 
a submanifold $P_f\hookrightarrow P_0$
(when it exists) where there are Hamiltonian vector fields $X_{H_0}\in\vf(P_0)$,
tangent to $P_f$, solutions to the contact Hamiltonian equations on $P_f$.
These vector fields solution are not unique, in general.
As in the Lagrangian formalism,
the constraint algorithm and the final dynamics
are independent of the Reeb vector field selected in the precontact manifold
$(P_0,\eta_0)$  (see \cite{DeLeon2019}).
\\ \indent
It can be proved that $P_f = \Leg(S_f)$,
where $S_f\hookrightarrow\Tan^{2k-1}Q$ is the submanifold where
there are vector field solutions to the higher-order Lagrangian equation \eqref{eq-E-L-contact1}
which are tangent to $S_f$ (see the above section).
Furthermore, as $\Leg_0$ is a submersion onto $P_0$, 
for every vector field $X_{H_0} \in \vf(P_0)$
which is a solution to the Hamilton equation \eqref{Hameqs0} on $P_f$, and tangent to $P_f$,
there exists some holonomic vector field $X_L \in \vf(\Tan^{2k-1}Q\times\R)$
which is a solution to the contact Euler-Lagrange equation \eqref{eq-E-L-contact1}
on $S_f$, and tangent to $S_f$,
such that ${\Leg_0}_*X_L= X_{H_0}$.
This $\Leg_0$-projectable holonomic vector field could be defined,
 in general, only on the
points of another submanifold $M_f\hookrightarrow S_f$.
(See \cite{art:Gracia_Pons_Roman91,art:Gracia_Pons_Roman92} 
for a detailed exposition of all these topics
in the case of higher-order presymplectic systems).
}\end{remark}


\section{Higher-order unified formalism}
\protect\label{uf}

The standard Lagrangian and Hamiltonian formalisms of higher-order contact systems can be recovered
from a single unified geometric framework which is a simpler and elegant way to present this theory.
This formulation is an extension of the one stated in \cite{LGMMR-2020}
for first-order contact systems and is inspired in the framework presented
in \cite{{art:Prieto_Roman11},{art:Prieto_Roman12}}.

\subsection{Unified bundle: precontact canonical structure}

Let $(\Tan^{2k-1}Q\times\R,L)$ be a (pre)contact Lagrangian system (of order $k$).

\begin{definition}
We define the \textbf{higher-order extended unified bundle}
(or \textbf{higher-order  extended Pontryagin bundle})
$$
\W=\Tan^{2k-1}Q\times_Q\Tan^*(\Tan^{k-1}Q)\times\R \ ,
$$
which is endowed with the natural submersions
$$
 \varrho_1\colon\W\to\Tan^{2k-1}Q\times\R \ ,\
 \varrho_2\colon\W\to\Tan^*(\Tan^{k-1}Q)\times\R \ ,\
\varrho_0\colon\W\to Q\times\R \ ,\ 
z\colon\W\to \Real \ .
\label{project}
$$
\end{definition}

The natural coordinates in $\W$ are $(q^i_0,\ldots,q^i_{2k-1},p_i^0,\ldots,p_i^{k-1},z)$.

\begin{definition}
A curve $\mbox{\boldmath $\sigma$}\colon\R\rightarrow\W$
is \textbf{holonomic} in $\W$ if the path 
$\varrho_1\circ\mbox{\boldmath $\sigma$}\colon\R\to\Tan^{2k-1}Q\times\R$ is holonomic.

A vector field $\Gamma \in\vf(\W)$  is a \textbf{semispray of order $1$}, 
or also a \textbf{holonomic vector field} in $\W$,
if its integral curves  are holonomic in $\W$. 
\end{definition}

In coordinates, a holonomic curve in $\W$ is expressed as 
$$
\mbox{\boldmath $\sigma$}=
\Big(q_0^i(t),\frac{dq_0^i}{d t}(t),\ldots,\frac{dq_{2k-1}^i}{d t}(t),p^0_i(t),\ldots,p^{k-1}_i(t),z(t)\Big) \ ,
$$
and a holonomic vector field in $\W$ is written as
\begin{equation}
\Gamma= 
q_1^i\derpar{}{q_0^i}+q_2^i\derpar{}{q_1^i}+\ldots
+q_{2k-1}^i\derpar{}{q_{2k-2}^i}+ f_{2k-1}^i\derpar{}{q_{2k-1}^i}
+g^0_1\derpar{}{p^0_i}+\ldots + g^{k-1}_i\derpar{}{p^{k-1}_i}+ 
g\,\frac{\partial}{\partial z} \ .
\label{holovfW}
\end{equation}

\begin{definition}
The bundle $\W$ is endowed with the following canonical structures:
\ben
\item
The \textbf{coupling function} is the
map ${\cal C}\colon\W\to\R$  defined as follows: 
let $w=({\rm p},\xi_q,z)\in\W$, where 
${\rm p}\in \Tan^{2k-1}Q$, $q = \rho^{2k-1}_{k-1}({\rm p})$ is its projection to $\Tan^{k-1}Q$, 
and $\xi_q \in \Tan_q^*(\Tan^{k-1}Q)$ is a covector; then
\begin{equation}\label{eqn:Cap06_DefCouplingFunc}
\begin{array}{rcl} {\cal C} \colon \Tan^{2k-1}Q \times_{\Tan^{k-1}Q} \Tan^*(\Tan^{k-1}Q) 
& \longrightarrow & \R \\ ({\rm p},\xi_q,z) & \longmapsto & \langle \xi_q \mid j_{k}({\rm p})_q \rangle \end{array} \ ,
\end{equation}
where $j_{k} \colon \Tan^{2k-1}Q \to \Tan(\Tan^{k-1}Q)$
is the canonical injection introduced in Section \ref{higherbundles},
 $j_{k}({\rm p})_q$ is the corresponding tangent vector to $\Tan^{k-1}Q$ in $q$,
 and $\langle \xi_q \mid j_{k}({\rm p})_q \rangle \equiv \xi_q(j_{k}({\rm p})_q)$
 denotes the canonical pairing between vectors of  $\Tan_q(\Tan^{k-1}Q)$ and
 covectors of $\Tan^*_q(\Tan^{k-1}Q)$.
\item
The \textbf{canonical $1$-form}
is the $\varrho_0$-semibasic form
$\Theta:=\varrho_2^*\,\theta_{k-1}\in\df^1(\W)$.
\\
The \textbf{canonical $2$-form} is
$\Omega:=-\d\Theta=\varrho_2^*\,\omega_{k-1}\in\df^2(\W)$.
\item
The \textbf{canonical precontact $1$-form}
is the $\varrho_1$-semibasic form
$\eta_{_\W}:=\varrho_2^*\eta_{k-1}=\d z-\Theta\in\df^1(\W)$.
\een
\label{coupling}
\end{definition}

In natural coordinates of $\W$ we have that
$$
\eta_{_\W}=\d z- \sum_{\alpha=0}^{k-1} p^\alpha_i\d q_\alpha^i \quad ,\quad
\d\eta_{_\W}=\sum_{\alpha=0}^{k-1}\d q_\alpha^i\wedge\d p^\alpha_i=\Omega 
\quad ,\quad
{\cal C}=\sum_{\alpha=0}^{k-1}p_i^\alpha q_{\alpha+1}^i
\ .
$$

\begin{definition}
Given a Lagrangian function $L\in\Cinfty(\Tan^kQ\times\R)$, if
$\Lag=\varrho_1^*L\in\Cinfty(\W)$,
we define the \textbf{Hamiltonian function} as
$$
\H:={\cal C}-\Lag=
\sum_{\alpha=0}^{k-1} p^\alpha_i q^i_{\alpha+1}-\Lag(q^i_\alpha,q^i_k,z)\in\Cinfty(\W) \ .
$$
\end{definition}

\begin{remark}{\rm
As $\eta_{_\W}$ is a precontact form in $\W$,
then $(\W,\eta_{_\W})$ is a precontact manifold and
$(\W,\eta_{_\W},\H)$ is a precontact Hamiltonian system.
As a consequence, the Reeb vector fields are not uniquely defined
but, as we have pointed out above, 
the formalism is independent of the choice of the Reeb vector field.
In this case, as $\W=\Tan^{2k-1}Q\times_Q\Tan^*(\Tan^{2k-1}Q)\times\R$
is a trivial bundle over $\R$, we can take the canonical lift to $\W$
of the canonical vector field $\displaystyle \derpar{}{z}$ of $\R$
as a representative of the family of Reeb vector fields.}
\end{remark}

\subsection{Higher-order contact dynamical equations}
\protect\label{des}

\begin{definition}
The \textbf{Lagrangian--Hamiltonian problem} for the contact system
$(\W,\eta_{_\W},{\cal H})$ consists in finding a vector field $X_\H\in\vf(\W)$
which is a solution to the contact Hamiltonian equations
\begin{equation}
\label{Whamilton-contact-eqs}
\inn(X_\H)\d\eta_{_\W}=\d\H-({\Reeb}(\H))\eta_{_\W}
\quad ,\quad
\inn(X_\H)\eta_{_\W}=-\H \ ,
\end{equation}
and then, the integral curves $\mbox{\boldmath $\sigma$}\colon I\subset\R\to\W$ of $X_\H$
are the solutions to equations
    \begin{equation}\label{Whamilton-contactc-curves-eqs}
\inn(\mbox{\boldmath $\sigma$}')\d\eta_{_\W}=
\left(\d\H-({\Reeb}(\H))\eta_{_\W}\right)\circ\mbox{\boldmath $\sigma$}
\quad ,\quad
 \inn(\mbox{\boldmath $\sigma$}')\eta_{_\W}=-\H\circ\mbox{\boldmath $\sigma$} \ ,
    \end{equation}
where $\mbox{\boldmath $\sigma'$}\colon I\subset\R\to \Tan\W$ is the canonical lift of
$\mbox{\boldmath $\sigma$}$ to $\Tan\W$.
\end{definition}

As $(\W,\eta_{_\W},\H)$ is a precontact Hamiltonian system,
these equations are not compatible in $\W$,
and we have to implement the constraint algorithm
in order to find the final constraint submanifold of $\W$
where there are consistent solutions to the equations.
So,  in a natural chart in $\W$, the local expression of a vector field is
\begin{equation}
X_\H = \sum_{\alpha=0}^{k-1}  
\left(f^i_\alpha\derpar{}{q^i_\alpha}+f_{k+\alpha}^i\derpar{}{q_{k+\alpha}^i}
+G^\alpha_i\derpar{}{p^\alpha_i}\right)+g\derpar{}{z} \in\vf(\W) \ ,
\label{coorvf}
\end{equation}
and therefore we obtain that
$$
\inn(X_\H)\eta_{_\W}=g-\sum_{\alpha=0}^{k-1}f^i_\alpha p_i^\alpha \quad , \quad
\inn(X_\H)\d\eta_{_\W}=\sum_{\alpha=0}^{k-1}(f^i_\alpha \,\d p_i^\alpha-G_i^\alpha\,\d q^i_\alpha) \ .
$$
Furthermore,
\begin{eqnarray*}
\d\H&=&
\sum_{\alpha=0}^{k-1} \left(q^i_{\alpha+1}\,\d p^\alpha_i+
\Big(p_i^\alpha -\derpar{\Lag}{q^i_{\alpha+1}}\Big)\d q^i_{\alpha+1}\right)  
-\derpar{\Lag}{q^i_0}\,\d q^i_0-\derpar{\Lag}{z}\,\d z \ , \\
({\Reeb}(\H))\eta_{_\W}&=&-\derpar{\Lag}{z}\Big(\d z-\sum_{\alpha=0}^{k-1}p_i^\alpha\d q^i_\alpha\Big) \ .
\end{eqnarray*}
The second equation \eqref{Whamilton-contact-eqs} gives
\begin{equation}
g=(f^i_\alpha-q^i_{\alpha+1})\,p_i^\alpha+\Lag \ ,
\label{zero}
\end{equation}
and the first equation \eqref{Whamilton-contact-eqs} leads to:
\begin{eqnarray}
 f_\alpha^i= q_{\alpha+1}^i \ , \label{one} \\
p^{k-1}_i= \derpar{\Lag}{q_k^i} \ ,  \label{two} \\
 G^0_i= \displaystyle \derpar{\Lag}{q_0^i}+
 p_i^0\derpar{\Lag}{z} \quad , \quad
G^r_i= -p^{r-1}_i+\displaystyle \derpar{\Lag}{q_r^i} +
 p_i^r\derpar{\Lag}{z}  \ .  \label{three}
\end{eqnarray}
where $0\leq\alpha\leq k-1$, $1 \leqslant r \leqslant k-1$,
and we have used the relation \eqref{recursivep} in the second group of equations \eqref{three}. Therefore:
\bit
\item
Equations (\ref{one}) are the conditions assuring that $X_\H$ is a semispray of type $k$ in $\W$,
regardless the regularity of the Lagrangian function.
This condition arises straightforwardly from the unified formalism
but, unlike what happens in order $k=1$, 
the unified formalism does not guarantee that $X_\H$ is a holonomic vector field:
it is holonomic up to order $k$ only.
\item
Using (\ref{one}), equation (\ref{zero}) leads to 
\begin{equation}
\label{g}
g=\Lag \ .
\end{equation}
\item
The algebraic equations (\ref{two}) are compatibility conditions;
that is, constraints defining locally a submanifold $\W_0\hookrightarrow\W$
where the vector fields $X_\H$ solution to \eqref{Whamilton-contact-eqs} are defined.
\item
Equations \eqref{three} allow us to determine 
all the component functions $G_i^\alpha$ in \eqref{coorvf}.
\eit
The component functions $f_{\alpha+k}^i$ in \eqref{coorvf} are still undetermined.
Nevertheless, the variational formulation demands the solution
to the dynamical equations to be holonomic curves,
which implies these vector fields $X_\H$ to be semisprays of type 1 in $\W$.
We can assure it by simply choosing the arbitrary functions
\begin{equation}
f_{\alpha+k}^i=q_{\alpha+k+1}^i \quad , \quad (0\leq\alpha\leq k-2) \ .
\label{holof}
\end{equation}
Thus, assuming holonomy, the
vector fields solution to \eqref{Whamilton-contact-eqs} are of the form
\begin{eqnarray}
\label{Xcoor}
X_\H &=& \sum_{\alpha=0}^{k-1}\left(q_{\alpha+1}^i\derpar{}{q_\alpha^i}+
q_{\alpha+k+1}^i\derpar{}{q_{\alpha+k}^i}\right)+ f^i_{2k-1}\derpar{}{q_{2k-1}^i}
\nonumber \\ & & 
\left(\derpar{\Lag}{q_0^i}+p_i^0\derpar{\Lag}{z} \right)\derpar{}{p^0_i}+
\sum_{r=1}^{k-1}\left(-p^{r-1}_i+\displaystyle \derpar{\Lag}{q_r^i} +
 p_i^r\derpar{\Lag}{z} \right)\derpar{}{p^r_i}+\L\derpar{}{z}
 \quad \ \mbox{\rm (on $\W_0$)} \ .
\end{eqnarray}
(In the appendix \ref{alternative} an alternative to this assumption,
which leads to an equivalent development of the constraint algorithm, is discussed).

At this point, the constraint algorithm continues by demanding that
$X_\H$ is tangent to $\W_0$; that is,
we have to impose that \
$\ds 
{X_\H\Big(p^{k-1}_i- \derpar{\Lag}{q_k^i}\Big)}\Big\vert_{_{\W_0}} 
= 0,$
then giving the constraints:
\begin{equation}
p^{k-2}_i=\displaystyle \derpar{\Lag}{q_{k-1}^i} +
 \derpar{\Lag}{q^i_k}\derpar{\Lag}{z}-\sum_{\alpha=0}^{k-1}q^j_{\alpha+1}\frac{\partial^2\Lag }{\partial q_\alpha^j\partial q_k^i}-\Lag\frac{\partial^2\Lag }{\partial z\partial q_k^i}\equiv
\displaystyle \derpar{\Lag}{q_{k-1}^i} - D_L\Big(\derpar{\Lag}{q^i_k}\Big)
 \ ;
\label{four}
\end{equation}
where the operator $D_L$ is given in \eqref{DL_operator}, thus recovering it also from the constraint algorithm.
Now the algorithm continues by demanding the tangency condition
on this new constraint submanifold, and repeating the process iteratively $k-2$ times, we get 
\begin{equation}
p^r_i=\derpar{\Lag}{q_{r+1}^i}-D_L(p^{r+1}_i) \quad , \quad
(0\leq r\leq k-3) \ .
\label{five}
\end{equation}
The constraints \eqref{two},  \eqref{four}, and \eqref{five} specify completely
the generalized momenta as derivatives of the Lagrangian $\L$,
and thus define completely the {\sl generalized Legendre map}.
Geometrically, these relations say that the vector fields $X_\H$ 
are defined on a submanifold $\W_1\hookrightarrow\W$
which is the graph of the extended Legendre map. 
$$
\W_1=\{ ({\rm p},\Leg({\rm p}))\in\W\,\mid\,{\rm p}\in\Tan^{2k-1}Q\}
={\rm graph}(\Leg) \ .
$$
Imposing once again the tangency condition on the constraints
obtained in the last step of the algorithm,
\ $\ds 
{X_\H\Big(p^0_i-\derpar{\Lag}{q_1^i}+D_L(p^1_i) \Big)}\Big\vert_{_{\W_1}} 
=~0$,
we obtain the equations for the remaining coefficients $ f^i_{2k-1}$ in \eqref{Xcoor}
which are
\begin{equation}
(-1)^k\Big(f_{2k-1}^j-D_L (q_{2k-1}^j)\Big) \frac{\partial^2L}{\partial q_k^j\partial q_k^i}+ \sum_{\alpha=0}^{k} (-1)^\alpha D_L^\alpha\Big( \derpar{L}{q_\alpha^i}\Big)=0 \quad (\text{on } \mathcal{W}_1)  \, .
\label{feq}
\end{equation}

Now, if $\mbox{\boldmath $\sigma$}(t)=(q^i_0(t),\ldots,q^i_{2k-1}(t),p_i^0(t),\ldots,p_i^{k-1}(t),z(t))$ 
is an integral curve of $X_\H$, equations 
\eqref{one}, \eqref{three}, \eqref{g}, \eqref{holof}, and \eqref{feq}
lead to the coordinate expression of equations
\eqref{Whamilton-contactc-curves-eqs} which is (on $\W_1$)
\begin{equation}
\begin{cases}
\ds \frac{d q_\alpha^i}{dt}&= q_{\alpha+1}^i \quad , \quad  (0 \leqslant\alpha \leqslant 2k-2) \ , \\
 0&=\ds (-1)^k\Big(\frac{d q_{2k-1}^j}{dt}-D_L (q_{2k-1}^j)\Big) \frac{\partial^2L}{\partial q_k^j\partial q_k^i}+ \sum_{\alpha=0}^{k} (-1)^\alpha D_L^\alpha\Big( \derpar{L}{q_\alpha^i}\Big)
 \ ,  \\
\ds \frac{d p^0_i}{dt}&= \displaystyle \derpar{\Lag}{q_0^i}+p_i^0\derpar{\Lag}{z} \ ,
\\
\ds \frac{d p^r_i}{dt} &= \displaystyle -p^{r-1}_i+\displaystyle \derpar{\Lag}{q_r^i} +
 p_i^r\derpar{\Lag}{z} 
\ , \quad (1\leq r\leq k-1) \ ,
\\
\ds \frac{d z}{dt} &=\L \ .
\end{cases}
\label{hoHeqs}
\end{equation}

\begin{remark}{\rm
If $L$ is regular, equations \eqref{feq} are compatible and define a unique vector field $X_\H$ solution to \eqref{Whamilton-contact-eqs} on $\W_1$,
and the last system of equations give the dynamical trajectories.
If $L$ is singular, equations \eqref{feq} can be compatible or not. 
Eventually, new constraints can appear and, in the most favourable cases,
there is a submanifold $\W_f \hookrightarrow \W_1$ (it could be $\W_f = \W_1$)
such that there exist holonomic vector fields $X_\H\in\vf(\W)$ defined on $\W_1$ and tangent to $\W_f$,
which are solutions to equations \eqref{Whamilton-contact-eqs}
at support on $\W_f$.}
\end{remark}

\begin{remark}{\rm
Summarizing, the unified formalism assures holonomy up to order $k-1$
and gives the higher-order momenta from the Lagrangian function.
Then, assuming holonomy up to the maximal order $2k-1$,
the constraint algorithm gives completely
the extended Legendre map, the final dynamical equations,
and also the operator $D_L$.}
\end{remark}

\subsection{Recovering the Lagrangian and Hamiltonian formalisms}
\label{recovering}

The Lagrangian and the Hamiltonian formalisms can be recovered from
the unified formalism using the natural projections 
and the extended Legendre map which has been obtained
as a consequence of the constraint algorithm.
The procedure follows the same pattern as for the unified formalism of first-order contact dynamical systems \cite{LGMMR-2020}.
(An exhaustive analysis on the equivalence among the Lagrangian, 
the Hamiltonian and the unified formalisms
for higher-order dynamical systems without dissipation
is made in \cite{{art:Prieto_Roman11}} and \cite{{art:Prieto_Roman12}}).

Let $P_0=\Leg(\Tan^{2k-1}Q\times\R)$, which is a submanifold of 
$\Tan^*(\Tan^{k-1}Q)\times\R$ when $L$ is an almost-regular Lagrangian and $P_0=\Tan^*(\Tan^{k-1}Q)\times\R$ when $L$ is hyperregular
(or an open set of $\Tan^*(\Tan^{k-1}Q)\times\R$ if $L$ is regular).
If we denote by  $\jmath_1\colon{\cal W}_1\hookrightarrow{\cal W}$
the natural embedding, we have that
$$
(\varrho_1\circ\jmath_1)({\cal W}_1)=\Tan^{2k-1}Q\times\R
\quad , \quad
(\varrho_2\circ\jmath_1)({\cal W}_1)=P_1\subseteq\Tan^*(\Tan^{k-1}Q)\times\R \ .
$$
As ${\cal W}_1={\rm graph}(\Leg)$,
it is diffeomorphic to $\Tan^{2k-1}Q\times\R$; that is,
the restricted projection $\rho_1\circ\jmath_1$ is a diffeomorphism.
Analogously, in the almost-regular case, for every submanifold
$\jmath_\iota\colon{\cal W}_\iota\hookrightarrow{\cal W}$
obtained from the constraint algorithm we have
$$
(\varrho_1\circ\jmath_\iota)({\cal W}_\iota)=S_\iota\hookrightarrow\Tan^{2k-1}Q\times\R
\quad , \quad
(\varrho_2\circ\jmath_\iota)({\cal W}_\iota)=P_\iota\hookrightarrow P_0\hookrightarrow\Tan^*(\Tan^{k-1}Q)\times\R \ ,
$$
and $\Leg(S_\iota)=P_\iota$, since 
${\cal W}_\iota\subseteq{\cal W}_1={\rm graph}(\Leg)$.
In particular, for the final constraint submanifold,
 $\jmath_f\colon{\cal W}_f\hookrightarrow{\cal W}_1$, we have that
$(\rho_1\circ\jmath_f)({\cal W}_f)=S_f\hookrightarrow\Tan^{2k-1}Q\times\R$
and
$(\rho_2\circ\jmath_f)({\cal W}_f)=P_f\hookrightarrow\Tan^*(\Tan Q)\times\R$
are the corresponding final constraint submanifolds of the Lagrangian and Hamiltonian constraint algorithms, respectively.
The equivalence between the constraint algorithms 
in the unified and in the Lagrangian formalism
only holds when the holonomy condition is also imposed
for the solutions in the Lagrangian case.
So we have the diagram
$$
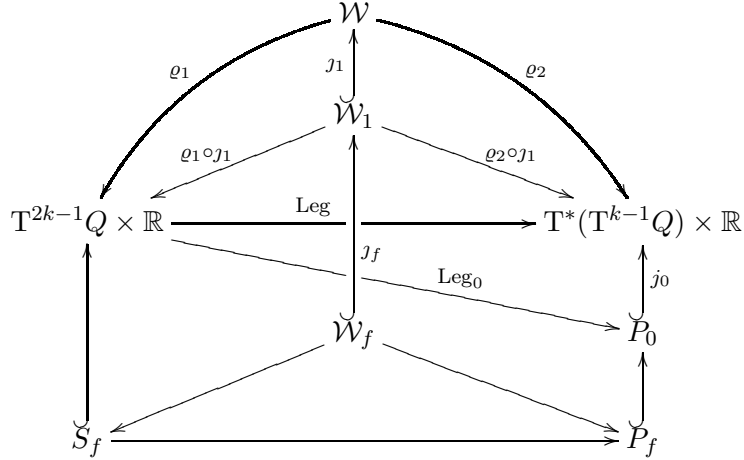
\xymatrix{
\ & \ & {\cal W} \ar@/_1.3pc/[ddll]_{\varrho_1} \ar@/^1.3pc/[ddrr]^{\varrho_2} & \ & \ \\
\ & \ & {\cal W}_1 \ar[dll]_{\varrho_1\circ\jmath_1} \ar[drr]^{\varrho_2\circ\jmath_1} \ar@{^{(}->}[u]^{\jmath_1} & \ & \ \\
\Tan^{2k-1}Q\times\R \ar[rrrr]^<(0.30){\Leg}|(.48){\hole}
\ar[drrrr]^<(0.60){\Leg_0}|(.480){\hole} 
& \ & \ & \ & \Tan^*(\Tan^{k-1}Q)\times\R  \\
\ & \ &  {\cal W}_f \ar@{^{(}->}[uu]_<(0.30){\jmath_f} \ar[dll]\ar[drr]  & \ & P_0 \ar@{^{(}->}[u]_{j_0} \\
S_f \ar@{^{(}->}[uu] \ar[rrrr] & \ & \ & \ & P_f \ar@{^{(}->}[u] \\
}
$$
Functions and differential forms in ${\cal W}$
and vector fields in ${\cal W}$ tangent to ${\cal W}_1$
can be restricted to ${\cal W}_1$. 
Then, they can be translated to the Lagrangian or the Hamiltonian side 
by using the diffeomorphism $\rho_1\circ\jmath_1$,
or projecting to $\Tan^*(\Tan^{k-1}Q)\times\R$.
In particular, observe that
$\Leg=(\varrho_2\circ\jmath_1)\circ(\varrho_1\circ\jmath_1)^{-1}$, and
$$
\theta_L=\Leg^*\theta_{k-1} \quad , \quad
\omega_L=\Leg^*\omega_{k-1} \quad , \quad
\eta_L=\Leg^*\eta_{k-1} \ .
$$
Therefore, the results and the discussion in the above section allow us to state:

\begin{teor}
\label{eqW}
Every curve $\mbox{\boldmath $\sigma$}\colon I\subseteq\Real\to{\cal W}$,
taking values in ${\cal W}_0$, can be written as
$\mbox{\boldmath $\sigma$}=(\mbox{\boldmath $\sigma$}_L,\mbox{\boldmath $\sigma$}_H)$, where
$\mbox{\boldmath $\sigma$}_L=\varrho_1\circ\mbox{\boldmath $\sigma$}\colon I\subseteq\Real \to\Tan^{2k-1}Q\times\R$
and $\mbox{\boldmath $\sigma$}_H=
\Leg\circ\mbox{\boldmath $\sigma$}_L\colon I\subseteq\Real\to P_0\subseteq\Tan^*(\Tan^{k-1}Q)\times\R$.

If $\mbox{\boldmath $\sigma$}\colon\Real\to{\cal W}$,
with  ${\rm Im}\,(\mbox{\boldmath $\sigma$})\subset{\cal W}_1$,
is a curve fulfilling equations \eqref{Whamilton-contactc-curves-eqs}
(at least on the points of a submanifold
${\cal W}_f \hookrightarrow{\cal W}_1$),
then
$\mbox{\boldmath $\sigma$}_L$ is the prolongation to
$\Tan^{2k-1}Q\times\R$ of the projected curve
${\bf c}=\varrho_0\circ\mbox{\boldmath $\sigma$}\colon\Real\to Q\times\R$ 
(that is, $\mbox{\boldmath $\sigma$}_L$ is a holonomic curve),
and it is a solution to equations \eqref{hec}.
Moreover, the curve  
$\mbox{\boldmath $\sigma$}_H=\Leg\circ\mbox{\boldmath $\sigma$}_L$
is a solution to equations \eqref{hamilton-contactc-curves-eqs} (on $P_f)$.

Conversely, if ${\bf c}\colon\Real\to Q\times\R$ is a curve such that
${\bf \bar c}^k=\mbox{\boldmath $\sigma$}_L$ is a solution to equation \eqref{hec} 
(on $S_f$), then the curve
$\mbox{\boldmath $\sigma$}=(\mbox{\boldmath $\sigma$}_L,\Leg\circ\mbox{\boldmath $\sigma$}_L)$
is a solution to equations \eqref{Whamilton-contactc-curves-eqs}.
In addition, $\Leg\circ\mbox{\boldmath $\sigma$}_L$
is a solution to equation \eqref{hamilton-contactc-curves-eqs} 
 \label{mainteor1} (on $P_f$).

(If $L$ is an almost-regular Lagrangian, then these results
hold on the points of the submanifolds ${\cal W}_f$,
$S_f$ and $P_f$).
\end{teor}

In coordinates, 
$\mbox{\boldmath $\sigma$}(t)=(q^i_0(t),\ldots,q^i_{2k-1}(t),p_i^0(t),\ldots,p_i^{k-1}(t),z(t))$ 
and therefore we have that
$\mbox{\boldmath $\sigma$}_L(t)=(q^i_0(t),\ldots,q^i_{2k-1}(t),z(t))$
and $\mbox{\boldmath $\sigma$}_H(t)=(q^i_0(t),\ldots,q^i_{k-1}(t),p_i^0(t),\ldots,p_i^{k-1}(t),z(t))$. Then, from
equations \eqref{hoHeqs} and using the extended Legendre map, 
we obtain that the components of $\mbox{\boldmath $\sigma$}_L(t)$
are the solutions to the system
\begin{align*}
\ds \frac{d q_\alpha^i}{dt}&= q_{\alpha+1}^i \quad , \quad  (0 \leqslant\alpha \leqslant 2k-2) \ , \\
  0&=(-1)^k\Big(\frac{d q_{2k-1}^j}{dt}-D_L (q_{2k-1}^j)\Big) \frac{\partial^2L}{\partial q_k^j\partial q_k^i}+ \sum_{\alpha=0}^{k} (-1)^\alpha D_L^\alpha\Big( \derpar{L}{q_\alpha^i}\Big)
 \ ,  \\
\ds \frac{d z}{dt} &=\L \ .
\end{align*}
or, what is equivalent, to the higher-order Euler-Lagrange or Herglotz equations  \eqref{ELeqs2} and \eqref{ELeqs1}.
On its turn, for the components of $\mbox{\boldmath $\sigma$}_H(t)$, from  \eqref{hoHeqs} 
(and using again the extended Legendre map when needed), we have
\begin{align*}
\ds \frac{d p^0_i}{dt}&= \displaystyle \derpar{\Lag}{q_0^i}+p_i^0\derpar{\Lag}{z} \ ,
\\
\ds \frac{d p^r_i}{dt} &= \displaystyle -p^{r-1}_i+\displaystyle \derpar{\Lag}{q_r^i} +
 p_i^r\derpar{\Lag}{z} \ , \quad (1\leq r\leq k-1) \ ,
\\
\ds \frac{d z}{dt} &=\L \ .
\end{align*}
which are the last group of Hamilton's equations \eqref{Heqcoor}.
The first group of Hamilton's equations \eqref{Heqcoor} arises
straightforwardly deriving the expression of the Hamiltonian function
(with respect to the momentum coordinates) and
 taking into account the holonomy conditions.

As the curves $\mbox{\boldmath $\sigma$}\colon\Real\to{\cal W}$ 
solution to equation \eqref{Whamilton-contactc-curves-eqs}
are the integral curves of holonomic vector fields $X_\H\in\vf({\cal W})$ 
solution to \eqref{Whamilton-contact-eqs}, and the curves 
$\mbox{\boldmath $\sigma$}_L\colon\Real\to\Tan^{2k-1}Q\times\R$
are the integral curves of holonomic vector fields 
$X_L\in\vf(\Tan^{2k-1}Q\times\R)$ 
solution to  \eqref{hec}, then as an immediate corollary of the above theorem 
we obtain:

\begin{teor}
\label{eqL}
Let $X_\H\in\vf({\cal W})$ be a solution to equations \eqref{Whamilton-contact-eqs}
(at least on the points of a submanifold
${\cal W}_f \hookrightarrow{\cal W}_1$) and tangent to 
${\cal W}_1$ (resp. tangent to ${\cal W}_f$). Then
the vector field $X_L\in\vf(\Tan Q\times\R)$, defined by
$X_L\circ\varrho_1=\Tan\varrho_1\circ X_\H$,
is a holonomic vector field (tangent to $S_f$)
solution to equations \eqref{eq-E-L-contact1} (on $S_f$),
where $E_L\in\Cinfty(\Tan^{2k-1}Q\times\R)$ is such that $\H=\varrho_1^*E_L$.

Furthermore, every holonomic vector field solution 
to equations \eqref{eq-E-L-contact1} (on $S_f$) 
can be recovered in this way from a vector field
$X_\H\in\vf({\cal W})$ (tangent to  ${\cal W}_f$)
solution to equations \eqref{Whamilton-contact-eqs}(on ${\cal W}_f$).
\end{teor}

The Hamiltonian formalism is recovered in a similar way,
taking into account that, now, the curves
$\mbox{\boldmath $\sigma$}_H\colon\Real\to\Tan^*Q\times\R$
are the integral curves of vector fields $X_H\in\vf(\Tan^*Q\times\R)$ 
solution to \eqref{hamilton-contact-eqs}.
So, for the regular case we have:

\begin{teor}
\label{eqH}
Let $X_\H\in\vf({\cal W})$ be a vector field, tangent to ${\cal W}_1$, which is
solution to equations \eqref{Whamilton-contact-eqs} (on $\W_1$).
Then the vector field $X_H\in\vf(\Tan^*(\Tan^{k-1}Q)\times\R)$, defined by
$X_H\circ\varrho_2=\Tan\varrho_2\circ X_\H$,
is a solution to equations \eqref{hamilton-contact-eqs},
where $H\in\Cinfty(\Tan^*(\Tan^{k-1}Q)\times\R)$ is such that $\H=\varrho_1^*H$.
This vector field satisfies that $\Leg_*X_L=X_H$.
\end{teor}

\begin{remark}{\rm
In the almost-regular case, 
the vector field $X_\H\in\vf({\cal W})$ is a
solution to equations \eqref{Whamilton-contact-eqs},
at least on the points of a submanifold
${\cal W}_f \hookrightarrow{\cal W}_1$, and is tangent to ${\cal W}_f$. 
Therefore, there exists a vector field $X_{H_0}\in\vf(P_0)$ which
is a solution to equations \eqref{hamilton-contact-eqs} in $P_0$
(with $H\in\Cinfty(P_0)$ such that $\H=(\Leg_0\circ\varrho_1)^*H$),
at least on the points of a submanifold $P_f\hookrightarrow P_0$, 
and is tangent to $P_f$.
This vector field verifies that
$(j_{o*}X_{H_0})\circ\varrho_2=\Tan\varrho_2\circ X_\H$.
}
\end{remark}


\section{Examples}
\protect\label{uex}

\subsection{The damped Pais--Uhlenbeck oscillator}

The Pais--Uhlenbeck oscillator is a typical example of a higher-order (regular) dynamical system, 
and has been analyzed in detail in many publications
(see, for instance, \cite{art:Martinez_Montemayor_Urrutia11,art:Pais_Uhlenbeck50}).

The configuration space for this system is a $1$-dimensional smooth manifold $Q$
with local coordinate $(q_0)$. Taking natural coordinates
in the higher-order tangent bundles over $Q$, 
the second-order Lagrangian function for this system is
$\ds L_0(q_0,q_1,q_2) = \frac{1}{2}( q_1^2 - \omega^2q_0^2 - \lambda q_2^2)  \in\Cinfty(\Tan^2Q)$;
where $\lambda$ is a nonzero real constant, and $\omega$ is another real constant.
It is a regular Lagrangian function because its Hessian matrix with respect to $q_2$ is
$\ds\left( \frac{\partial^2L_0}{\partial q_2\partial q_2} \right) =- \lambda$,
which has maximum rank since $\lambda\not=0$
(if $\lambda = 0$, then $L_0$ becomes a first-order regular Lagrangian).
The corresponding $4$\,th-order Euler--Lagrange equation is
\beq
\frac{d^4q_0}{dt^4}=- \frac{1}{\lambda}\left(\omega^2q_0+\frac{d^2q_0}{dt^2}\right) \ , 
\label{PUos}
\eeq

We can introduce damping in this model by adding a standard dissipation term
to the Lagrangian as follows
$$
L=L_0-\gamma  z= \frac{1}{2} \left( q_1^2 - \omega^2q_0^2 - \lambda q_2^2 \right)-\gamma z\in\Cinfty(\Tan^2Q\times\R) \quad ; \quad \gamma\in\R^+ \ .
$$
Next, we study it using the unified formalism.
The unified phase space of the system is
$ \Tan^3Q \times_{\Tan Q} \Tan^*(\Tan Q)\times\R$
and we have the following diagram
$$
\xymatrix{
\ & \W = \Tan^3Q \times_{\Tan Q} \Tan^*(\Tan Q)\times\R \ar[dl]_-{\varrho_1} \ar[dr]^-{\varrho_2} & \ \\
\Tan^3Q\times\R \ar[r]_-{\rho^{3}_{1}\times{\rm Id}} & \Tan Q\times\R & \Tan^*(\Tan Q)\times\R \ar[l]^-{\pi_{\Tan Q}\times{\rm Id}} \ .
}
$$
As usual, we denote by $\L\in\Cinfty(\W)$ the pullback of $L$ to $\W$
(which has the same coordinate expression than $L$).
If $\vartheta_1 \in \df^1(\Tan^*(\Tan Q))$ is the canonical $1$-form,
we define the canonical pre-contact structure
$$
\eta_{_\W}=\d z-(\pi_1\circ\varrho_2)^*\vartheta_1=
\d z- p^0\,\d q_0 - p^1\, \d q_1 \in \df^1(\W) \ ,
$$
and the  local expression of the Hamiltonian function
$\H={\cal C}-\varrho_1^*L={\cal C} -\L\in\Cinfty(\W)$ is
\begin{equation*}
\H(q_0,q_1,q_2,q_3,p^0,p^1,z) = p^0q_1 + p^1q_2 - \frac{1}{2} \left( q_1^2 - \omega^2q_0^2 - \lambda q_2^2 \right)+\gamma z \ .
\end{equation*}
A generic vector field $X_\H\in\vf(\W)$ is locally given by
$$
X_\H=f_0\derpar{}{q_0}+f_1 \derpar{}{q_1}+F_2 \derpar{}{q_2}+F_3\derpar{}{q_3} +
 G^0\derpar{}{p^0} + G^1\derpar{}{p^1}+g\derpar{}{z} \ ;
$$
then, taking into account that
$$
\d\H = \omega^2q_0\d q_0 + (p^0-q_1)\d q_1 + (p^1 + \lambda q_2)\d q_2 + q_1 \d p^0 + q_2 \d p^1+\gamma\d z \ ,
$$
from the contact dynamical equations \eqref{Whamilton-contact-eqs} we obtain \begin{align}
& g=(f_0-q_1)p^0+(f_1-q_2)p^1+\L \ ,
\label{1} \\
 f_0 = q_1 \quad , \quad
 f_1 = q_2 \quad , \quad &
 p^1 + \lambda q_2 = 0 \quad ,\quad
 G^0 = - \omega^2 q_0-\gamma p^0 \quad , \quad
 G^1 = q_1 - p^0-\gamma p^1 \ .
\label{2}
\end{align}
The first two equations in \eqref{2}
give us the condition of semispray of type $2$ for the vector field $X_\H$,
and, going to \eqref{1}, it implies that $g=\L$.
The third equation in \eqref{2}
is a constraint defining the submanifold $\W_0$ where $X_\H$ exists.
If we require $X_\H$ to be a semispray of type $1$, then $F_2=q_3$
and the local expression of $X_\H$, on $\W_0$,  is
$$
X_\H = q_1 \derpar{}{q_0} + q_2 \derpar{}{q_1} + q_3 \derpar{}{q_2} + F_3 \derpar{}{q_3} -
 (\omega^2q_0+\gamma p^0)\derpar{}{p^0} +
 \left(q_1 - p^0-\gamma p^1\right) \derpar{}{p_1}+\L\derpar{}{z} \ .
$$
The tangency condition for $X_\H$ on $\W_0$ gives the new constraint
$$
0=\Lie(X_\H)\xi_0\equiv X_\H(p^1+\lambda q_2)= q_1 - p^0-\gamma p^1+\lambda q_3\equiv\xi_1
\quad \ \mbox{\rm (on $\W_0$)}  \ ,
$$
which defines the submanifold
$\W_1 = \left\{ p \in \W\,\vert\, \xi_0(p) = \xi_1(p) = 0 \right\}$,
that we identify with the graph of the extended Legendre map 
$\Leg \colon \Tan^3Q\times\R \to \Tan^*(\Tan Q)\times\R$
which is given as
$$
\Leg^*p^1 = \derpar{\L}{q_2} = - \lambda q_2  \quad , \quad
\Leg^*p^0 =
\derpar{\L}{q_1} -D_L\left(\derpar{\L}{q_2}\right) \equiv
 \derpar{\L}{q_1} -D_L\left( p^1 \right) = 
q_1 + \lambda q_3-\gamma p^1  \ ,
$$
and the identity on the other coordinates $(q_0,q_1,z)$.
As $\lambda \not= 0$, we see that $L$ is a regular Lagrangian since
$\Leg$ is a (local) diffeomorphism. Observe that
$$
 D_L(F)=q_1\frac{\partial F}{\partial q_0}+q_2\frac{\partial F}{\partial q_1}+
q_3\frac{\partial F}{\partial q_2}+\Lag\frac{\partial F}{\partial z} + \gamma  F\ .
$$
Next we compute the tangency condition for $X_\H$ on $\W_1$ which leads to
$$
0= \Lie(X_\H)\xi_1=X_\H(q_1 - p^0-\gamma p^1+\lambda q_3)=
\lambda F_3+q_2+\omega^2q_0+2\gamma p^0-\gamma \left(q_1-\gamma p^1\right)
\quad \ \mbox{\rm (on $\W_1$)}  \ ,
$$
and this equation allows us to determine $F_3$  univocally.
Thus, there is a unique vector field $X_\H$ solution to equations \eqref{Whamilton-contact-eqs},
which is tangent to $\W_1$, and is given locally by
\begin{eqnarray*}
X_\H&=& q_1 \derpar{}{q_0} + q_2 \derpar{}{q_1} + q_3 \derpar{}{q_2} + 
\frac{1}{\lambda}\left(-\omega^2q_0+\gamma q_1-q_2-2\gamma p^0-\gamma^2 p^1\right) \derpar{}{q_3} \\ & &
-(\omega^2q_0+\gamma p^0)\derpar{}{p^0} +
 \left(q_1 - p^0-\gamma p^1\right) \derpar{}{p_1}+\L\derpar{}{z} \ .
\end{eqnarray*}
Now, if $\mbox{\boldmath $\sigma$}\colon\R\to\W$, with
$\mbox{\boldmath $\sigma$}(t)=(q_0(t),q_1(t),q_2(t),q_3(t),p^0(t),p^1(t),z(t))$,
is an integral curve of $X_\H$, its component functions are solutions
 to the system differential equations 
\begin{eqnarray}
 \dot{q}_0(t) = q_1(t)  &,& 
 \dot{q}_1(t) = q_2(t) \ , 
\label{first} \\
 \dot{q}_2(t) = q_3(t)  &,& 
 \dot{q}_3(t) = \frac{1}{\lambda}\left(\gamma q_1(t)-\omega^2q_0(t)-q_2(t)-2\gamma p^0(t)-\gamma^2p^1(t)\right)\ , \qquad
\label{second} \\
 \dot{p}^0(t) =  - \left(\omega^2q_0(t)+\gamma p^0(t)\right)  &,& 
 \dot{p}^1(t) = q_1(t) - p^0(t)-\gamma p^1(t)  \ ,
\label{third} \\
& & \dot z=\frac{1}{2} \left( q_1(t)^2 - \omega^2q_0(t)^2 - \lambda q_2(t)^2 \right)-\gamma z(t) \ .
\label{fourth}
\end{eqnarray}
The final diagram is
$$
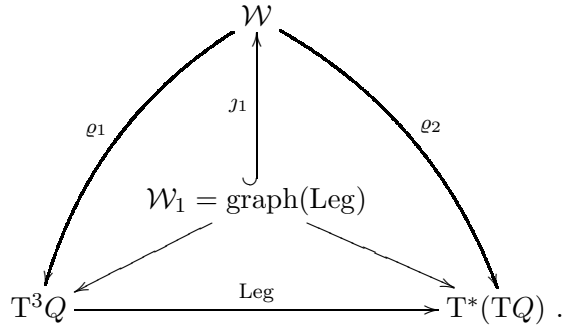
\xymatrix{
\ & \W \ar@/_1.25pc/[dddl]_-{\varrho_1} \ar@/^1.25pc/[dddr]^-{\varrho_2} & \ \\
\ & \ & \ \\
\ & \W_1 = {\rm graph}(\Leg) \ar@{^{(}->}[uu]^-{\jmath_1} \ar[dl]_-
{} \ar[dr]^-{} & \ \\
\Tan^3Q \ar@{->}[rr]^-{\Leg} & \ & \Tan^*(\Tan Q) \ .
}
$$

Finally, from the unified formalism, we can recover
the Lagrangian and Hamiltonian solutions for this system
using the natural projections and the constraints defining $\W_1$; that is, the extended Legendre map.

For the Lagrangian formalism, as we
have shown in Theorem \ref{eqL}, the Euler-Lagrange vector field
solution to \eqref{eq-E-L-contact1} is the unique
semispray of type $1$, $X_L \in \vf(\Tan^3Q)$, such that
$X_L \circ \varrho_1= \Tan \varrho_1 \circ X_\H$,
and it is locally given by
$$
X_L= q_1 \derpar{}{q_0} + q_2 \derpar{}{q_1} + q_3 \derpar{}{q_2} - 
\frac{1}{\lambda}\left(\omega^2q_0+\gamma q_1+(1+\lambda\gamma^2)q_2+2\gamma\lambda q_3\right) \derpar{}{q_3}+\L\derpar{}{z}  \ .
$$
For the integral curves of $X_L$
we have that, if $\mbox{\boldmath $\sigma$}(t)=(q_0(t),q_1(t),q_2(t),q_3(t),p^0(t),p^1(t),z(t))$ is an integral curve of $X_\H$, 
then,  from Theorem \ref{eqW},
$\mbox{\boldmath $\sigma$}_L= \varrho_1\circ\mbox{\boldmath $\sigma$}$
is an integral curve of $X_L$
and its components $\mbox{\boldmath $\sigma$}_\Lag(t)=(q_0(t),q_1(t),q_2(t),q_3(t),z(t))$
are solutions to equations \eqref{first} and \eqref{second} which read as
\begin{eqnarray*}
 \dot{q}_0(t) = q_1(t) \ &,& \
 \dot{q}_1(t) = q_2(t) \ , 
\\
 \dot{q}_2(t) = q_3(t)  \ &,& \ 
 \dot{q}_3(t) =-\frac{1}{\lambda}\left(\omega^2q_0(t)+\gamma q_1(t)+(1+\lambda\gamma^2)q_2(t)+2\gamma\lambda q_3(t)\right) \ , 
\\ & &  \dot z=\frac{1}{2} \left( q_1(t)^2 - \omega^2q_0(t)^2 - \lambda q_2(t)^2 \right)-\gamma z(t) \ .
\end{eqnarray*}
and combining the first four equations we arrive to
$$
\frac{d^4q_0}{dt^4}=-\frac{1}{\lambda}\left(\omega^2q_0(t)+\gamma\frac{dq_0}{dt}+(1+\lambda\gamma^2)\frac{d^2q_0}{dt^2}+2\gamma\lambda\frac{d^3q_0}{dt^3}\right)\ , 
$$
which is the Euler--Lagrange equation for the damped Pais-Uhlenbeck oscillator.
Observe that, if $\gamma=0$, we recover
the dynamical equation for
the usual Pais--Uhlenbeck oscillator \eqref{PUos}.

For the Hamiltonian formalism, as $L$ is a regular Lagrangian,
Theorem \ref{eqH} states that there exists a unique
vector field $X_H=\Leg_*X_L\in \vf(\Tan^*(\Tan Q))$ which is solution to
\eqref{Hameqs}, and, hence, it is locally given by
\begin{eqnarray*}
X_H&=& q_1 \derpar{}{q_0} -\frac{p^1(t)}{\lambda} \derpar{}{q_1} 
-(\omega^2q_0+\gamma p^0)\derpar{}{p^0} +
 \left(q_1 - p^0-\gamma p^1\right) \derpar{}{p_1}
\\ & &+
\left(\frac{1}{2} \Big(q_1^2-\omega^2q_0^2-\frac{(p^1)^2}{\lambda}\Big)-\gamma z\right)\derpar{}{z}  \ .
\end{eqnarray*}
For its integral curves,  Theorem \ref{eqW} states that
if $\mbox{\boldmath $\sigma$}_L\colon \R \to \Tan^3Q$ 
is an integral curve of $X_L$ related with
an integral curve $\mbox{\boldmath $\sigma$}$ of $X_\H$,
then $\mbox{\boldmath $\sigma$}_H= \Leg\circ\mbox{\boldmath $\sigma$}_L$
is an integral curve of $X_h$. Therefore, if 
$\mbox{\boldmath $\sigma$}_H(t) =(q_0(t),q_1(t),p^0(t),p^1(t),z(t))$,
its components are solutions to equations \eqref{first}, \eqref{third}, and \eqref{fourth} (with $\ds q_2(t)=-\frac{p^1(t)}{\lambda})$,
which are the standard Hamilton equations for this system.

\subsection{Integrating factors: the radiating electron}

A common technique to model a differential equation in a symplectic setting consists in using  an integrating factor \cite{Ra-2006}. This is a non-vanishing function (or positive-definite matrix) which appears multiplying the ODE in the final Euler-Lagrange equations. When a Lagrangian does not require an integrating factor it is called a \emph{universal Lagrangian} \cite{Cawley-1979}.

We will focus in the case of Lagrangians with the form 
$\hat{L}=\exp(-\gamma t) l(q^i_0,q_1^i,\dots,q_k^i)$. 
In this case the Euler-Lagrange equations are proportional to $\exp(-\gamma t)$.
This is a usual integrating factor when modelling systems with dissipation. 
It turns out that the ODE described by these Lagrangians can be obtained in the contact setting by the Lagrangian $L=l(q^i_0,q_1^i,\dots,q_k^i)-\gamma z$.
Indeed, the higher-order Euler--Lagrange equations for the symplectic case \eqref{eq:HOsymplectic} are
$$
\sum_{\alpha=0}^{k}(-1)^\alpha d_T^\alpha\Big(\derpar{\hat{L}}{q_{\alpha}^i}\Big) = 0 \ ,
$$
which have the same structure as the higher-order contact Euler--Lagrange \eqref{ELeqs2}, but changing the operator $d_T$ by $D_{L}$.  For the symplectic case, we have:
$$
d_T(\exp(-\gamma t) f(q^i_0,q_1^i,\dots,q_k^i))=\exp(-\gamma t)\left(\sum_{\alpha=0}^k q^j_{\alpha+1}\frac{\partial f}{\partial q^j_\alpha}-\gamma f\right) \ ;
$$
meanwhile, for the contact case,
$$
D_{L}f(q^i_0,q_1^i,\dots,q_k^i)=\sum_{\alpha=0}^k q^j_{\alpha+1}\frac{\partial f}{\partial q^j_\alpha}-\gamma f \ .
$$
Therefore
$$
\sum_{\alpha=0}^{k}(-1)^\alpha d_T\Big(\derpar{\hat{L}}{q_{\alpha}^i}\Big) = \exp(-\gamma t)\sum_{\alpha=0}^{k}(-1)^\alpha D_L^\alpha\Big(\derpar{L}{q_{\alpha}^i}\Big) \ .
$$

Although both formalisms lead to the same dynamical system, they have different geometric structures. The contact approach directly produces equations, thus $L$ is a sort of universal Lagrangian for the contact setting. Both uses an auxiliary variable  to include dissipation. In the symplectic case, the time is used, leading to a non-autonomous system and cosymplectic geometry. On the contact case, the auxiliary variable is $z$. Cosymplectic structures are the natural description for non-autonomous systems \cite{DeLeon2016b}, while contact geometry describe action-dependent Lagrangians, in particular some cases of dissipation. For these reasons, we argue that contact geometry is more adequate for this kind of systems.

As an example, consider the motion of an electron in a potential $V(q_0^i)$, using a non-relativistic setting but taking into account electrodynamic reaction forces. It can be modelled by \cite{art:Caldirola,art:Valentini}
$$
m\frac{\tau^2}{16}q_4^i-m\frac{\tau}{2}q_3^i+mq_2^i=-\frac{\partial V}{\partial q_0^i}\quad ;\quad i=1,2,3  \ ,
$$
where $\tau$ is a non-vanishing real constant. This equation can be obtained with the symplectic formalism from the  Lagrangian
 \cite{art:Caldirola}
$$
L'=-\exp(-\frac{4}{\tau}t)\Big(m\frac{\tau^2}{32}\sum_{i=1}^3(q_2^i)^2+V\Big) \ ,
$$
which uses a non-autonomous integrating factor. It can also be obtained using the higher-order contact formalism presented here from the following second-order regular Lagrangian
$$
L=m\frac{\tau^2}{32}\sum_{i=1}^3(q_2^i)^2+V+\frac{4}{\tau}z \ .
$$

\subsection{Singular case: $\gamma az$ term}

The standard dissipation term added in a mechanical Lagrangian is $\gamma z$, and it is the one used in the previous examples. In \cite{GGMRR-2019b} a term proportional to velocity times $z$ is used to obtain a dissipation quadratic in velocities. 
Here we consider a term proportional to the acceleration times $z$.

The configuration space for this system is a $1$-dimensional smooth manifold $Q$
with local coordinate $(q_0)$. Taking natural coordinates
in the higher-order tangent bundles over $Q$, 
we consider the Lagrangian
$\ds L(q_0,q_1,q_2) = \frac{1}{2} m q_1^2 -V(q_0)-\gamma q_2 z \in\Cinfty(\Tan^2 Q\times \mathbb{R})$;
where $m$ is a positive real constant and $\gamma$ is a real constant with units of time over distance.  This is a singular second-order Lagrangian. The corresponding contact Euler-Lagrange equation is
\begin{equation}
\label{sing:E-L}
mq_2\left(\frac12\gamma^2q_1^2+2\gamma q_1-\frac{\gamma^2}{m}V+1\right)+(1-\gamma q_1)\frac{\partial V}{\partial q_0}=0 \ ,
\end{equation}
which is a second-order ODE, although we expect it to be 4th-order. This is due to the degeneracy of the Lagrangian with respect to accelerations (see \cite{GR-2016} for a study on order reduction in higher-order symplectic mechanics). 
Now, we apply the unified formalism for this Lagrangian, 
showing how to compute it for singular Lagrangians.

The unified formalism takes place on the manifold $\mathcal{W}=T^3Q\times_{TQ}T^*(TQ)\times \mathbb{R}$, with natural coordinates $(q_0,q_1,q_2,q_3,p^0,p^1,z)$. It is the same space as the one from the damped Pais--Uhlenbeck oscillator example, and the reader can look there the geometrical details. Over this manifold, we consider the precontact structure given by $\eta=\d z-p^0\d q_1-p^1\d q_2$, and pick as Reeb vector field $\ds R=\frac{\partial}{\partial z}$. The pullback of $L$ to $\mathcal{W}$ will be denoted $\mathcal{L}$, and has the same local expression. The corresponding Hamiltonian is
$$
\mathcal{H}=p^0q_1+p^1q_2-\frac12mq_1^2+V(q_0)+\gamma q_2 z\in C^\infty(\mathcal{W}) \ .
$$
A generic vector field $X_\H\in\vf(\W)$ is locally given by
$$
X_\H=f_0\derpar{}{q_0}+f_1 \derpar{}{q_1}+F_2 \derpar{}{q_2}+F_3\derpar{}{q_3} +
 G^0\derpar{}{p^0} + G^1\derpar{}{p^1}+g\derpar{}{z} \ .
$$
With these elements, the contact dynamical equations \eqref{Whamilton-contact-eqs} become
\begin{align}
f_0 = q_1 \quad , \quad
 f_1 = q_2 \quad , \quad  g=(f_0-q_1)p^0+(f_1-q_2)p^1+\L \ ,
\label{sing1} \\
 p^1 + \gamma z = 0 \quad ,  \quad 
 G^0 = -V'-\gamma q_2 p^0  \quad , \quad
 G^1 = mq_1-p^0-\gamma q_2 p^1 \ .
\label{sing2}
\end{align}
In \eqref{sing1} we partially recover the holonomy of $X_\H$, and derive that $g=\mathcal{L}$, as expected. The first equation in \eqref{sing2} is a constraint that defines the momentum $p^1$. 
The tangency condition on this constraint leads to the second constraint
$$
-\gamma\mathcal{L}=mq_1-p^0-\gamma q_2 p^1\quad\Longleftrightarrow\quad p^0=\frac{\gamma}{2}mq_1^2-\gamma V+mq_1 \ ;
$$
which determines the momentum $p^0$. The tangency condition on it,
$$
\Lie(X_\mathcal{H})\Big(p^0-\frac{\gamma}{2}mq_1^2-\gamma V+mq_1\Big)=0 \ ,
$$
results in the Euler-Lagrange equations \eqref{sing:E-L}.

With the unified algorithm we recover the Legendre transformation from the constraint algorithm. 
The Euler-Lagrange equation is derived, but they appear as a constraint on the manifold.
Furthermore, not all the holonomy is recovered, and $F_2$ and $F_3$ are not determined. 
This is due to the projectability of the Lagrangian to a lower-order tangent bundle. 
If we impose the whole holonomy, $F_2=q_3$, the tangency condition allows us to determine $F_3$,
$$
F_3=\frac{-m\gamma^2q_1q_2^2-2m\gamma q_2^2+\gamma^2q_1q_2\frac{\partial V}{\partial q_0}+\gamma q_2\frac{\partial V}{\partial q_0}-(q_1-\gamma q_1^2)\frac{\partial^2 V}{\partial q_0^2}}{\frac12m\gamma^2q_1^2+2m\gamma q_1-\gamma^2V+m} \ .
$$


\section{Conclusion and outlook}
\protect\label{di}

The standard Lagrangian formalism for systems described
by contact Lagrangian functions of order $k$ takes place in the
{\sl extended higher-order tangent bundle} $\Tan^{2k-1}Q\times\R$.
First, we have stated the corresponding variational formulation,
which is a generalization of Herglotz's variational principle
for first-order contact Lagrangians,
and we obtain the dynamical equations for these kinds of systems.
Then, in order to construct the geometric setting for this theory,
we need to use the canonical geometric structures 
of higher-order tangent bundles and the basic ideas of contact mechanics.
As a previous relevant step, the concept of
total derivative in $\Tan^kQ$, introduced by Tulczyjew,
had to be generalized in $\Tan^{2k-1}Q\times\R$.
It is noticed that this extension (which is defined using
certain kinds of holonomic vector fields in $\Tan^{2k-1}Q\times\R$)
is not canonical because it depends on the Lagrangian function $L$.
With this operator, we can define a contact
(or precontact) structure in  $\Tan^{2k-1}Q\times\R$, 
the {\sl contact Lagrangian form} (which depends on $L$), 
and allows us to state the geometric dynamical contact equations of the system.
Their solutions must be vector fields which are required to be 
semisprays of type 1 in that bundle and whose integral curves
are holonomic paths which are solutions to the higher-order 
contact Euler-Lagrange (Herglotz) equations
obtained from the higher-order Herglotz variational principle.
The requirement that vector fields solution must be holonomic
(semisprays of type 1) is a condition for the dynamical equations to be variational.
The contact Lagrangian form is also used to define the
Legendre map $\Leg\colon\Tan^{2k-1}Q\times\R\longrightarrow\Tan^*(\Tan^{k-1}Q)\times\R$
and then to establish the canonical Hamiltonian formalism 
for these higher-order contact Lagrangian systems.
This formalism is just a standard contact Hamiltonian formalism
adapted to the cotangent bundle $\Tan^*(\Tan^{k-1}Q)$.

A simpler and elegant equivalent formulation of the theory
is based in using an extension of the unified Lagrangian--Hamiltonian formalism
of Skinner and Rusk.
In this framework, neither the structures of  higher-order tangent bundles, nor
the extension of the Tulczyjew total derivative are needed, in principle.
In fact, the natural coupling between elements of the bundles 
$\Tan(\Tan^{k-1}Q)\times\R$ and $\Tan^*(\Tan^{k-1}Q)\times\R$
and the Lagrangian can be used to construct a Hamiltonian function which,
together with the natural contact form that $\Tan^*(\Tan^{k-1}Q)\times\R$
is endowed with, allows us to define a precontact Hamiltonian system
in the unified bundle $\W=\Tan^{2k-1}Q \times_Q\Tan^*(\Tan^{k-1}Q)\times\R$.
Being this system singular, the constraint algorithm must be implemented
to solve the dynamical equations. Then, if we look for holonomic solutions,
the algorithm leads to obtain the complete Legendre map
(the momentum coordinates are obtained as constraints),
the extension of the total derivative and both
the Euler--Lagrange (Herglotz) and the Hamilton equations,
thus recovering the standard Lagrangian and Hamiltonian formalisms.
It is interesting to point out that, alternatively,
we can seek for general solutions (non necessarily holonomic vector fields);
then the constraint algorithm fixes only partially the holonomy (up to order $k$)
and the Legendre map (giving only the higher-order momenta),
and the extended total derivative is then needed to complete the Legendre map,
but the rest of the holonomy conditions are then
given by the constraint algorithm (see the appendix \ref{alternative}).

As a first interesting example we have studied a regular system
which comes from extending the Lagrangian of the {\sl Pais--Uhlenbeck oscillator}
which introduces damping in the model.
The resulting final ($4$\,th-order) dynamical equation
incorporates additional terms which are responsible
for dissipation and that are proportional, not only to velocity
(as it is usual in first-order systems with damping) but also
to acceleration and its time-derivative (which sometimes is called ``jerk'' or `jolt'').
The second example is a modelling of a radiating electron
using a second-order Lagrangian which avoids to introduce the usual
``exponential factor'' in the standard first-order Lagrangian for the system.
A final academic example is a system described by a second order singular Lagrangian 
with a dissipation term of the type $\gamma a z$
which illustrates the behaviour of higher-order singular systems.

The results of this paper open many possibilities for future work.
One of the subjects to be investigated are the symmetries of contact higher-order mechanics; indeed,
the usual symplectic case has been offered a lot of interesting results by extending Noether's theorem \cite{LM-95,GP-95}. 
The main difference now is that in the contact setting, 
symmetries of the system produce dissipated quantities
\cite{DeLeon2019,DeLeon2019b,GGMRR-2019b}. 
In addition, the study of the momentum map is also relevant
for a better understanding of these systems.

Moreover,  we can study  the  extension to higher-order non-autonomous systems, 
where we have an explicit
dependence on time; so, we would need to extend the notion of contact geometry to obtain a proper  geometric setting. 

Another line of research is the study of the discrete version of the theory. 
This has a particular interest since it would provide a method to develop geometric integrators.

Of course, the extension to vakonomic systems and the applications to optimal control theory
are also part of our plan of research. 
Let us recall that some of us have recently obtained a {\sl Pontryagin
Maximum Principle} for contact optimal control theories \cite{LLM-2020}.

Finally, we want to explore the possibility to apply all the results in this paper and the new ones that we could obtain
to thermodynamical systems.

\appendix

\section{Alternative development of the constraint algorithm in the unified formalism}
\label{alternative}

As we have seen in Section \ref{des},
in the unified Lagrangian--Hamiltonian formalism
the use of the constraint algorithm plays a crucial role.
In that Section, after obtaining the compatibility conditions
\eqref{two} that define the submanifold $\W_0$ where $X_\H$ 
exist and are semisprays of type $k$,
the holonomy condition for the vector fields was imposed
before analysing the tangency condition.
Nevertheless, there is an alternative way to proceed which consists 
in, instead to impose the holonomy of $X_\H$, to
complete the generalized Legendre map using the recursive relations
\eqref{recursivep} as new constraints to define the 
submanifold $\W_1={\rm graph}(\Leg)\hookrightarrow\W$. 
In this way, the definition of the extended Legendre map 
is a consequence of the constraint algorithm and the relations \eqref{recursivep}. 
Then, the vector fields solution to \eqref{Whamilton-contact-eqs}
are of the form
\begin{eqnarray*}
X_\H &=& \sum_{\alpha=0}^{k-1}\left(q_{\alpha+1}^i\derpar{}{q_\alpha^i}+
f_{\alpha+k}^i\derpar{}{q_{\alpha+k}^i} \right)+ 
\left(\derpar{\Lag}{q_0^i}+p_i^0\derpar{\Lag}{z} \right)\derpar{}{p^0_i}
\\ & &+
\sum_{r=1}^{k-1}\left(-p^{r-1}_i+\displaystyle \derpar{\Lag}{q_r^i} +
 p_i^r\derpar{\Lag}{z} \right)\derpar{}{p^r_i}+\L\derpar{}{z}
 \ ,
\quad \mbox{\rm (on $\W_1$)} \ ,
\end{eqnarray*}
where the functions $f_{\alpha+k}^i$ are still undetermined.
The constraint algorithm continues by demanding that
$X_\H$ is tangent to $\W_1$; that is,
we have to impose that ${X_\H(\xi)}\vert_{_{\W_1}} = 0$,
for every constraint function $\xi$ defining $\W_1$.
Thus we obtain the following $kn$ equations (on $\W_1$)
\begin{eqnarray}
\label{TanVectFieldX}
0&=&\displaystyle \left(f_k^j-q_{k+1}^j\right)\derpars{\Lag}{q_k^j}{q_k^i} \ , \nonumber \\
0&=&\displaystyle \left(f_{k+1}^j - q_{k+2}^j\right)\derpars{\Lag}{q_k^j}{q_k^i} - 
\left(f_k^j-q_{k+1}^j \right) D_L\Big(\derpars{\Lag}{q_k^j}{q_k^i}\Big) \ ,  \nonumber \\
& & \qquad  \qquad  \qquad  \qquad  \qquad \vdots \\
0&=&\displaystyle \left(f_{2k-2}^j - q_{2k-1}^j\right)\derpars{\Lag}{q_k^j}{q_k^i} -
 \sum_{\alpha=0}^{k-3} \left(f_{k+\alpha}^j-q_{k+\alpha+1}^j\right) (\cdots\cdots) \ ,  \nonumber \\
0&=&\displaystyle (-1)^k\Big(f_{2k-1}^j-D_L(q_{2k-1}^j)\Big) \derpars{\Lag}{q_k^j}{q_k^i} + 
\sum_{\alpha=0}^{k} (-1)^\alpha D_L^\alpha\Big( \derpar{\Lag}{q_\alpha^i} \Big)
-\ds\sum_{\alpha=0}^{k-2} \Big(f_{k+\alpha}^j-q_{k+\alpha+1}^j\Big) (\cdots\cdots)  \nonumber \ ,
\end{eqnarray} 
where the terms in brackets $(\cdots\cdots)$ contain relations involving partial derivatives
of the Lagrangian and applications of $D_L$.
The compatibility of these equations depends on the regularity of the
Lagrangian function.  In particular:

\begin{prop}
\label{prop:Cap06_RegLag}
If $L \in \Cinfty(\Tan^kQ\times\R)$ is a regular Lagrangian function, then
there exists a unique vector field $X_\H \in \vf(\W)$ on $\W_1$,
which is a solution to equation \eqref{Whamilton-contact-eqs},
it  is  tangent to $\W_1$, and it is a semispray of type $1$ in $\W$.
\end{prop}
\proof
The regularity of the Hessian matrix
$\ds \left(\derpars{\Lag}{q_k^j}{q_k^i}\right)$ at every point
allows us to determine all the functions $f_\alpha^i$ uniquely
as the solution to the system of equations \eqref{TanVectFieldX},
which leads to \eqref{four} and \eqref{five}.
Therefore, the tangency condition holds for $X_\H$ on $\W_1$.
Furthermore, the equalities (\ref{four}) show that
$X_\H$ is a holonomic vector field in $\W$ (see the local expression \eqref{holovfW}).
\qed

\begin{remark}{\rm
If $L$ is singular, the system of $kn$ equations \eqref{TanVectFieldX} can be compatible or not. 
Eventually, new constraints can appear and, in the most favourable cases,
there is a submanifold $\W_f \hookrightarrow \W_1$ (it could be $\W_f = \W_1$)
such that there exist vector fields $X_\H\in\vf(\W)$ on $\W_1$ 
and tangent to $\W_f$, which are solutions to equations 
\eqref{Whamilton-contact-eqs} on $\W_f$.
These vector fields are semisprays of order $k$ but not necessarily holonomic 
in $\W$ and, in order to obtain holonomic vector fields solution,
the equalities \eqref{four} must be imposed ``ad hoc''.
Therefore, the tangency condition could originate new constraints which define
another submanifold ${\cal S}_f \hookrightarrow \W_f$ where 
there are holonomic vector fields $\Gamma_\H\in\vf(\W)$, 
defined on ${\cal S}_f$ and tangent to ${\cal S}_f$,
which are solutions to equations \eqref{Whamilton-contact-eqs} on ${\cal S}_f$.}
\end{remark}

\subsection*{Acknowledgments}

We acknowledge the financial support from the Spanish
Ministerio de Ciencia, Innovaci\'on y Universidades project
PGC2018-098265-B-C33,
the MINECO Grant MTM2016-76-072-P, 
the ICMAT Severo Ochoa projects SEV-2011-0087 and SEV-2015-0554,
and the Secretary of University and Research of the Ministry of Business and Knowledge of
the Catalan Government project 2017--SGR--932.
Manuel La\'inz wishes to thank MICINN and ICMAT for a FPI-Severo Ochoa predoctoral contract PRE2018-083203.


\addcontentsline{toc}{subsection}{\bf References}
\itemsep 0pt plus 1pt

\end{document}